\documentclass[aps,prd,twocolumn,floatfix,nofootinbib,showpacs,superscriptaddress,tightenlines]{revtex4}

\usepackage{amsmath}
\usepackage{dcolumn}
\usepackage{lipsum}
\usepackage{amssymb}
\usepackage{epsfig}
\usepackage{graphicx}
\usepackage{amsmath}
\usepackage{bm}
\usepackage{setspace}
\usepackage{lscape}
\usepackage{amsthm}
\usepackage{bbold}
\usepackage{dcolumn}
\usepackage{epsfig}
\usepackage{graphics}
\usepackage{graphicx}
\usepackage{longtable}
\usepackage{color}
\usepackage{xcolor}
\usepackage{bm}
\usepackage{xspace}
\usepackage{cancel}
\usepackage{float}
\usepackage{multirow}
\definecolor{darkgreen}{rgb}{0,0.5,0}
\definecolor{purple}{rgb}{0.5,0,0.5}
\definecolor{nblue}{rgb}{0.0,0.0,0.50}
\definecolor{scarlet}{rgb}{1.0,0.2,0}
\definecolor{darkmagenta}{rgb}{0.55, 0.0, 0.55}
\definecolor{darkolivegreen}{rgb}{0.33, 0.42, 0.18}
\definecolor{darkcandyapplered}{rgb}{0.64, 0.0, 0.0}
\usepackage[colorlinks=true, pdfstartview=FitV, linkcolor=purple, citecolor= purple, urlcolor=blue]{hyperref}
\usepackage{cleveref}
\newcommand{\be}{\begin{equation}}
\newcommand{\tu}{\textcolor{red}{u}}

\newcommand{\fu}{\textcolor{blue}{\bar{f_2}}}
\newcommand{\fd}{\textcolor{blue}{f_1}}
\newcommand{\fdu}{\textcolor{blue}{f_2}}

\newcommand{\Me}{\textcolor{blue}{V}}
\newcommand{\Meps}{\textcolor{blue}{{PS}}}

\newcommand{\Jpsi}{\textcolor{blue}{J/\Psi}}

\newcommand{\g}{\textcolor{blue}{\Gamma}}
\newcommand{\D}{\textcolor{magenta}{\Delta}}
\newcommand{\td}{\textcolor{darkcandyapplered}{d}}
\newcommand{\tb}{\textcolor{blue}{b}}
\newcommand{\tc}{\textcolor{darkmagenta}{c}}
\newcommand{\ts}{\textcolor{darkgreen}{s}}
\newcommand{\ee}{\end{equation}}
\newcommand{\bea}{\begin{eqnarray}}
\newcommand{\eea}{\end{eqnarray}}
\newcommand{\beas}{\begin{eqnarray*}}
\newcommand{\eeas}{\end{eqnarray*}}
\newcommand{\nn}{\nonumber}

\newcommand{\tq}{\textcolor{red}{q}}
\newcommand{\tqu}{\textcolor{blue}{q_1}}
\newcommand{\tdS}{\textcolor{blue}{SD}}
\newcommand{\tavd}{\textcolor{blue}{AVD}}
\newcommand{\MeV}{\text{MeV}} 
\newcommand{\GeV}{\text{GeV}} 
\newcommand{\rmh}{\hat{\alpha}_{\mathrm {IR}}}

\begin{document}
\title{Masses of Light and Heavy Mesons and Baryons: A Unified Picture}

\author{L. X. Guti\'{e}rrez-Guerrero}
\affiliation{
CONACyT-Mesoamerican Centre for Theoretical Physics, Universidad Aut\'onoma de Chiapas,
Carretera Zapata Km. 4, Real del Bosque (Ter\'an), Tuxtla Guti\'errez 29040, Chiapas,
M\'exico.}

\author{Adnan Bashir}
\affiliation{Instituto de F\'{\i}sica y Matem\'aticas, Universidad Michoacana de San Nicol\'as de Hidalgo,
Edificio C-3, Ciudad Universitaria, C.P. 58040, Morelia, Michoac\'an, M{\'e}xico}

\author{Marco A. Bedolla}
\affiliation{
Mesoamerican Centre for Theoretical Physics, Universidad Aut\'onoma de Chiapas,
Carretera Zapata Km. 4, Real del Bosque (Ter\'an), Tuxtla Guti\'errez 29040, Chiapas, M\'exico}
\affiliation{ Istituto Nazionale di Fisica Nucleare (INFN), Sezione di Genova, via Dodecaneso 33,
16146 Genova, Italia.}

\author{E. Santopinto}
\affiliation{ Istituto Nazionale di Fisica Nucleare (INFN), Sezione di Genova, via Dodecaneso 33,
16146 Genova, Italia.}

\begin{abstract}
 We compute masses of positive parity spin-$1/2$ and $3/2$ baryons composed of $\tu$, $\td$, $\ts$, $\tc$ and
 $\tb$ quarks in a quark-diaquark picture. The mathematical foundation for this analysis is implemented
 through a symmetry-preserving Schwinger-Dyson equations treatment of a vector-vector contact interaction,
 which preserves key features of quantum chromodynamics, such as confinement, chiral symmetry breaking and low energy Goldberger-Treiman relations.
 This study requires a computation of diquark correlations containing these quarks which in turn
 are readily inferred from solving the Bethe-Salpeter equations of the corresponding mesons. Therefore,
 it serves as a unified formalism for a multitude of mesons and baryons. It builds on our previous works
 on the study of masses, decay constants and form factors of quarkonia and light mesons,
 employing the same model.
 We use two sets of parameters, one which remains exactly the same for both the
 light and heavy sector hadrons, and another where the coupling strength is allowed to evolve according to
 the available mass scales of quarks. Our results are in very good agreement with the existing experimental
 data as well as predictions of other theoretical approaches whenever comparison is
 possible.
\end{abstract}

\pacs{12.38.-t, 12.40.Yx, 14.20.-c, 14.20.Gk, 14.40.-n, 14.40.Nd, 14.40.Pq}

\maketitle

\section{Introduction}

Comprehensive theoretical and experimental studies of baryons containing charm and bottom
quarks have been a focus of invigorated research over the last several years.
Recall that the dynamics of light quarks is dominated by the emergent
phenomena of dynamical chiral symmetry breaking (DCSB) and confinement, orchestrated by quantum
chromodynamics (QCD).
The presence of charm and/or bottom quarks with masses significantly greater
than $\Lambda_{\rm QCD} \approx 300$ MeV provides a flavor tag as it introduces a mass scale much
larger than the scale of confinement and effective light quark masses within hadrons.
Consequently, an exploration of such states presents an opportunity to quantitatively understand the
swapping roles played by the chiral dynamics of the light sector and the heavy quark symmetries
when charm and bottomom quarks are involved.
The quark model in the heavy quark sector predicts hadrons having one, two or three heavy quarks ($\tc,\tb$) as constituents~\cite{Roberts:2007ni,Crede:2013sze,Chen:2016spr}. Properties and decays of singly heavy baryons have been widely studied, see for example~\cite{Lu:2014ina,Majethiya:2009vx}. Note that the singly heavy
and light baryons can shed light on the role played by the diquark correlations between light quarks,
the non-pointlike dynamical degrees of freedom which are known to dictate baryon properties. In the case of light
baryons, such investigations cannot provide a comparative analysis of the relative contribution of light-light and heavy-light diquarks because all the diquark correlations inside them are made of light quarks and follow similar dynamics. In the presence of one charm or bottom quark, we expect to understand the quantitative relative contribution of different diquark correlations~\cite{Gottlieb:2007ay}. These light quark and heavy-light diquark correlations
reflect in the ground and excited state spectra, masses, decay rates, branching ratios and the production rates.

The excited spectra of doubly heavy baryons and their splittings can shed
light into their intrinsic collective degrees of freedom, which are characterized by two
widely separated scales : the low momentum scale of the light quark ($\sim \Lambda_{QCD}$) and
the relatively heavy charm or bottom quark mass, giving rise to excitation in these systems.
The triply heavy baryons are expected to be the ideal candidates for understanding
the QCD dynamics for such systems~\cite{Bjorken:1985ei,Wang:2018utj}.

There is a very active ongoing experimental program at various laboratories to study charm
and bottom baryons, their masses, lifetimes and weak decays.
BELLE has done a lot of spectroscopy of $\Xi_c$ and $\Omega_c$ resonances, measuring precisely the masses and widths of several such states. With 50 times more data obtained through upgrades to the KEK accelerator facility, Belle II will continue with charmed baryon spectroscopy. With data taking under way, it should precisely measure line-shapes, map out resonances, search for new decay channels
and test predictions for hitherto unobserved states~\cite{Schwartz:2017gni}.
Exciting QCD and hadron physics program at the Jefferson Laboratory unravel
the internal structure of ground and excited state mesons and baryons in terms of their
electromagnetic and tranition form factors~\cite{Aguilar:2019teb,Aznauryan:2012ba,Bashir:2012fs}.
Excited strange and charmed baryons will also be studied at future experiments like PANDA~\cite{Lin:2014jza}.
The LHCb detector is designed for the study of particles containing $\tb$ or $\tc$ quarks. They have recently reported two new baryons, named $\Sigma_{b}(6097)^{\pm}$ which appear as resonances in the two-body system $\Lambda_b^0 \pi^{\pm}$~\cite{Aaij:2018tnn}. These states can perhaps be identified as
$P$-wave excited states,~\cite{Chen:2018vuc}.
Especially interesting is the observation of a doubly charmed baryon, called
$\Xi_{\tc\tc}^{++}(\tu\tc\tc)$, seen by the LHCb collaboration in the $\Delta_c^+K^-\pi^+\pi^+$ final state, with mass around 3621 MeV~\cite{Aaij:2017ueg}. This value is higher than that for the first doubly charmed state $\Xi_{\tc\tc}^{++}(\tu\tc\tc)$ measured by SELEX in 2002. Its mass was determined to be 3460 MeV~\cite{Mattson:2002vu,Russ:2002bw}.

We shall present the calculation of heavy baryon masses using a continuum treatment of nonperturbative QCD, namely the Schwinger-Dyson equations (SDEs). These were earlier used to study the spectrum and interactions of mesons with masses less than 1 GeV~\cite{Maris:2003vk}.
The rainbow-ladder truncation was later applied to the ground state heavy-heavy mesons~\cite{Bhagwat:2006xi}.
A vector-vector contact-interaction (CI) was first proposed in~\cite{GutierrezGuerrero:2010md}. Since then it has been used to study a wide range of mesons in light quark sector~\cite{GutierrezGuerrero:2010md,Roberts:2010rn,Roberts:2011wy,Chen:2012qr,Xu:2015kta}.
Its range of applicability was later extended to the study of charmed and bottom mesons~\cite{Bedolla:2015mpa,Serna:2017nlr,Raya:2017ggu,Yin:2019bxe}. A calculation of the spectrum of strange and nonstrange hadrons with the CI treatment was published in~\cite{ Chen:2012qr}. Parity partners in the baryon resonance spectrum were later computed in~\cite{Lu:2017cln}. Masses of the ground-state mesons and baryons, including those containing heavy quarks were presented recently in~\cite{Yin:2019bxe,Chen:2019fzn}.
Note that heavy baryons have also been studied using numerical simulations of lattice-regularised QCD, see for example~\cite{Brown:2014ena,Padmanath:2015jea,Mathur:2018epb}. When we present our results of the CI, We make comparisons, whenever possible, with earlier representative studies in the field.


We organize this paper as follows. In section~\ref{CI-1}, we present our formulation of CI model. The Bethe-Salpeter (BS) equation to study two-particles bound states is introduced in section~\ref{BS-s}. We then calculate the diquarks and mesons containing charm and bottom quarks. Pseudoscalar mesons and their diquark partners are studied in section~\ref{Ps-mesons}. The masses of vector mesons and axial-vector diquarks are computed in section~\ref{mesons-V}.  In section~\ref{Baryons}, we describe the Faddeev equation (FE), focusing on baryons with one, two or three heavy quarks. The results for masses of spin $1/2$ and spin $3/2$ baryons are discussed in sections~\ref{baryons-1} and~\ref{baryons-2}, respectively. Finally, a summary and outlook is provided in section~\ref{Conclusions}.

\section{ Contact interaction: features }
\label{CI-1}
The gap equation for fermions requires modelling the gluon propagator and the quark-gluon vertex.
Here we shall recall and list these key characteristics of the CI~\cite{GutierrezGuerrero:2010md,Roberts:2010rn,Roberts:2011wy,Roberts:2011cf}~:
\begin{itemize}
\item  The gluon propagator is defined to be independent of any varying momentum scale:
\begin{eqnarray}
\label{eqn:contact_interaction}
g^{2}D_{\mu \nu}(k)&=&\frac{4\pi\alpha_{\mathrm{IR}}}{m_g^2}\delta_{\mu \nu} \equiv
\frac{1}{m_{G}^{2}}\delta_{\mu\nu},
\end{eqnarray}
\noindent where $m_g=500\,\MeV$ is a gluon mass scale generated dynamically in QCD~\cite{Boucaud:2011ug,Aguilar:2017dco,Binosi:2017rwj,Gao:2017uox}, and $\alpha_{\mathrm{IR}}$ can be interpreted as the interaction strength in the infrared~\cite{Binosi:2016nme,Deur:2016tte,Rodriguez-Quintero:2018wma}.
\item At  leading-order, the quark-gluon vertex is
\begin{equation}
\Gamma_{\nu}(p,q) =\gamma_{\nu}
\end{equation}
\item With this kernel the dressed-quark propagator for a quark of flavor $f$ becomes
\begin{eqnarray}
 \nn && S_f^{-1}(p) = \\
&&  i \gamma \cdot p + m_f +  \frac{16\pi}{3}\frac{\alpha_{\rm IR}}{m_G^2} \int\!\frac{d^4 q}{(2\pi)^4} \,
\gamma_{\mu} \, S_f(q) \, \gamma_{\mu}\,,\label{gap-1}
\end{eqnarray}
where $m_f$ is the current-quark mass. The integral possesses quadratic and logarithmic divergences and we regularize them in a Poincar\'e covariant manner to preserve the axial-vector Ward-Takahashi identity. The solution is~:
\begin{equation}
\label{genS}
S_f^{-1}(p) = i \gamma\cdot p + M_f\,,
\end{equation}
 where $M_f$ in general is a mass function running with a momentum scale, but within the CI it
 is a constant dressed mass.
\item $M_f$ is determined by
\begin{equation}
M_f = m_f + M_f\frac{4\alpha_{\rm IR}}{3\pi m_G^2}\,\,{\cal C}^{\rm iu}(M_f^2)\,,
\label{gapactual}
\end{equation}
where
\bea
\hspace{0.75 cm}
{\cal C}^{\rm iu}(\sigma)/\sigma = \overline{\cal C}^{\rm iu}(\sigma) = \Gamma(-1,\sigma \tau_{\rm uv}^2) - \Gamma(-1,\sigma \tau_{\rm ir}^2),
\eea
 with $\Gamma(\alpha,y)$ being the incomplete gamma-function and $\tau_{\rm ir, uv}$ are respectively, infrared and ultraviolet regulators. A nonzero value for  $\tau_{\mathrm{IR}}\equiv 1/\Lambda_{\mathrm{IR}}$ implements
confinement~\cite{Roberts:2007ji}. Since the CI is a nonrenormalizable theory,
$\tau_{\mathrm{UV}}\equiv 1/\Lambda_{\mathrm{UV}}$ becomes part of the model and therefore sets the scale for
all dimensional quantities.
 \end{itemize}
 We report results  using two parameters sets, the light-quark parameters (CI-LP), and the heavy ones (CI-HP), Table~\ref{parameters1}. The parameters denoted by CI-LP are used in calculations of heavy and light hadron masses unlike the CI-HP, which are a function of the mass of the constituent quarks.
 \begin{table}[htbp]
 \caption{\label{parameters1} Dimensionless coupling constant $\hat{\alpha}=\hat{\alpha}_{\mathrm IR}\Lambda_{\mathrm UV}^{2}$,
  where $\rmh=\alpha_{\mathrm {IR}}/m_{g}^{2}$, for the CI, extracted from a
  best-fit to data, as explained in Ref.~\cite{Raya:2017ggu}. Fixed parameters are $m_{g}=0.50\,\GeV$ and
  $\Lambda_{\mathrm {IR}}=0.24\,\GeV$.}
\begin{center}
\label{parameters1}
\begin{tabular}{@{\extracolsep{0.0 cm}}lcc}
\hline \hline
& Light-quark parameters (CI-LP)  & \\
 quarks & $\hat{\alpha}_{\mathrm {IR}}\;[\GeV^{-2}]$ & $\Lambda_{\mathrm {UV}}\;[\GeV] $  \\
$\tu,\td,\ts,\tc,\tb$ & 4.57 & 0.91 \\
\hline
& Heavy-quark parameters (CI-HP)& \\
 quarks & $\hat{\alpha}_{\mathrm {IR}}\;[\GeV^{-2}]$ & $\Lambda_{\mathrm {UV}}\;[\GeV] $  \\
$\tc,\td,\ts$ & 0.96 & 1.32 \\
$\tc$     & 0.22 & 2.31 \\
$\tb,\tu,\ts$ & 0.18 & 2.22 \\
$\tb,\tc$   & 0.05 & 4.24 \\
$\tb$     & 0.01 & 7.37 \\
\hline \hline
\end{tabular}
\end{center}
\end{table}
 %
 %
%
%
If one wants to go beyond predicting the masses of the hadrons and construct a model which can also predict
charge radii and decay constants, then the study of the heavy sector requires a change in the model parameters with respect to those of the light sector: an increase in the ultraviolet regulator, and a reduction in the coupling strength. Following Ref.~\cite{Raya:2017ggu}, guided by~\cite{Farias:2005cr,Farias:2006cs}, we define a dimensionless coupling $\hat{\alpha}$~:
\begin{equation}
\hat{\alpha}(\Lambda_{\mathrm{UV}})=\rmh\Lambda_{\mathrm{UV}}^2.\label{eqn:dimensionless_alpha}
\end{equation}
In close analogy with the running coupling of QCD with the momentum scale on which it is measured, an inverse logarithmic curve can describe the functional dependence of $\hat{\alpha}(\Lambda_{\mathrm{UV}})$ reasonably well:
\begin{equation}
\label{eqn:logaritmicfit} \hat{\alpha}(\Lambda_{\mathrm{UV}})=a\ln^{-1}\left(\Lambda_{\mathrm {UV}}/\Lambda_0\right) \,,
\end{equation}
$a=0.92$ and $\Lambda_0=0.36$ GeV, see Fig. 1 in Ref.~\cite{Raya:2017ggu}. With this expression, we can estimate the
value of the coupling strength $\hat{\alpha}(\Lambda_{\mathrm{UV}})$ once a value of $\Lambda_{\mathrm {UV}}$
is assigned.

 Table~\ref{table-M} presents the values of $\tu$, $\ts$, $\tc$ and $\tb$ dressed quark masses computed from Eq.~(\ref{gapactual}).
%
\begin{table}[htbp]
\caption{\label{table-M}
Computed dressed-quark masses (in \GeV), required as input for the BS equation and FE.}
\vspace{0.3cm}
\begin{tabular}{@{\extracolsep{0.2 cm}}cccc}
\hline
\hline
\rule{0ex}{2.0ex}
Light-quark  & \hspace{-0.7cm} parameters & \hspace{-1.5cm} (CI-LP)& \\
\hline
\rule{0ex}{2.0ex}
 $m_{\tu}=0.007$ &$m_{\ts}=0.17$ & $m_{\tc}=1.58$ & $m_{\tb}=4.83$   \\
  $\hspace{-.2cm} M_{\tu}=0.36$ & $ \hspace{0.1cm} M_{\ts}=0.53$\; &  $M_{\tc}=1.60$ & $M_{\tb}=4.83$  \\
\hline
Heavy-quark & \hspace{-0.7cm} parameters & \hspace{-1.5cm} (CI-HP) &  \\
\hline
 $m_{\tu}=0.007$ &$m_{\ts}=0.17$ & $m_{\tc}=1.08$ & $m_{\tb}=3.80$   \\
 $\hspace{-.2cm} M_{\tu}=0.36$ & $ \hspace{0.1cm} M_{\ts}=0.53$\; &  $ \hspace{0.0 cm} M_{\tc}=1.53$ & $ \hspace{0.0cm} M_{\tb}=4.68$  \\
 \hline
 \hline
\end{tabular}
\end{table}
The simplicity of the CI allows one to readily compute hadronic observables, such as masses, decay constants, charge radii and form factors. The study of heavy, heavy-light and light meson masses should provide a way to determine diquark effective masses, assumed to be confined within baryons, and properties of heavy, heavy-light, and light baryons. With this in mind, in the next section, we describe and solve the BS equation for mesons and diquarks.
\section{Masses of mesons and diquarks containing $\tc$ and $\tb$ quarks}
\label{BS-s}
The dominant correlations for ground state baryons are
scalar (0+) and axial-vector (1+) diquarks. At leading-order in a symmetry preserving truncation
of the SDEs, simple changes in the equations describing
 mesons yield expressions that provide
detailed information about the scalar and axial-vector diquarks.
Such states are of course not colorless
in QCD, and are therefore expected to be confined, if they exist at all. Nevertheless,
the masses of these states can serve as an indication for the relevant
mass scales of quark-quark correlations. Such diquark correlations could play a
role inside baryons: two quarks strongly correlated in a color antitriplet configuration can
couple with a quark to form a color-singlet baryon. With this purpose, a calculation of the
physical meson masses is a guide to compute its diquark partner masses.
%
%
The bound-state problem for hadrons characterized by two valence-fermions may be studied using the
homogeneous BS equation in Fig.~\ref{BSEfig}. This equation is~\cite{Salpeter:1951sz}
\begin{figure}[ht]
\vspace{-6cm}
       \centerline{
       \includegraphics[scale=0.5,angle=0]{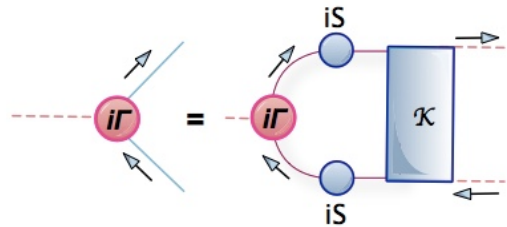}}
       \label{BSEfig}
       \vspace{-5.5cm}
       \label{BSE}\caption{This diagram represents the BS equation. Blue (solid) circles represent dressed propagators $S$, red (solid) circle is the meson BS amplitude $\Gamma$ and the blue (solid) rectangle is the dressed-quark-antiquark scattering kernel $K$.}
\end{figure}
\vspace{-0.5cm}
\begin{equation}
[\Gamma(k;P)]_{tu} = \int \! \frac{d^4q}{(2\pi)^4} [\chi(q;P)]_{sr} K_{tu}^{rs}(q,k;P)\,,
\label{genbse}
\end{equation}
where $\Gamma$ is the bound-state's BS amplitude; $\chi(q;P) = S(q+P)\Gamma S(q)$ is the BS wave-function; $r,s,t,u$ represent colour, flavor and spinor indices; and $K$ is the relevant fermion-fermion scattering kernel.  This equation possesses solutions on that discrete set of $P^2$-values for which bound-states exist.
We use the notation introduced in~\cite{Chen:2012qr}, [$\fd,\fdu$] for scalar diquarks, and $(\{\fd,\fd\}),(\{\fd,\fdu\})$ for axial vector  diquarks. We will describe the spectroscopy of
charm and bottom mesons in the following subsections.
\subsection{Pseudoscalar mesons and scalar diquarks}
\label{Ps-mesons}
\begin{figure}[htpb]
\vspace{-4.29cm}
       \centerline{\includegraphics[scale=0.55,angle=0]{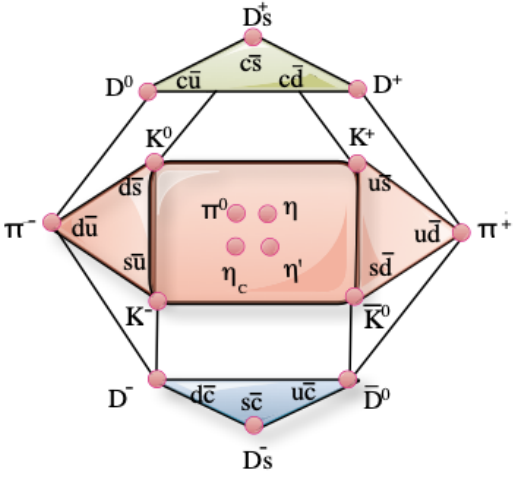}}
       \vspace{-4.5cm}
       \label{meson-2}\caption{The multiplet for the pseudoscalar mesons made of $\tu,\;\td,\;\ts$ and $\tc$ quarks.
        \vspace{-1.0cm}}
       \label{meson-pse}
\end{figure}

\begin{table*}[ht]
\caption{\label{table-mesones-pseudo}
Computed masses for pseudoscalar mesons and scalar diquarks (in \GeV) with the parameters in Table~\ref{parameters1}. Experimental masses are taken from~\cite{Patrignani:2016xqp, Tanabashi:2018oca}}
\begin{center}
\begin{tabular}{@{\extracolsep{0.5 cm}}cccccccc}
\hline
\hline
Mesons   & $\eta_c(\tc\bar{\tc})$ & $D^{0}(\tc\bar{\tu})$ & $D^{+}_{\ts}(\tc\bar{\ts})$  & $B_{\tc}^{+}(\tc\bar{\tb})$ &  $B^{+} (\tu\bar{\tb})$ &  $\eta_{\tb}(\tb\bar{\tb})$ & $B_s^0(\ts\bar{\tb})$ \\
Expt. & 2.98 &1.86&1.97 & 6.27 & 5.28&9.40& 5.37\\
CI-LP & 2.98 & 1.82   & 1.94 &6.19&5.07 & 9.46 & 5.17\\
CI-HP & 2.97 & 1.86   & 1.95 &6.28& 5.28 & 9.44 & 5.37\\
\hline
Diquarks & $ [\tc\tc]_{0+}$ & $ [\tc\tu]_{0+}$& $ [\tc\ts]_{0+}$& $ [\tc\tb]_{0+}$ &  $ [\tu\tb]_{0+}$ &  $ [\tb\tb]_{0+}$  & $[\ts\tb]_{0+}$\\
CI-LP &3.11& 2.01 & 2.13 &6.31 & 5.23& 9.53 & 5.34 \\
CI-HP &3.17 & 2.07 & 2.17 &6.33 & 5.36& 9.43 & 5.42 \\
\hline
\hline
\end{tabular}
\end{center}
\end{table*}
We consider hadrons with five quark flavors ($\tu$,$\td$,$\ts$,$\tc$,$\tb$) with SU(5) multiplets predicted
by the quark model~\cite{GellMann:1964nj,Zweig:1964jf,Zweig:1981pd}. As an example, a set of corresponding pseudoscalar mesons projected along the ($\tu$,$\td$,$\ts$,$\tc$)-axes,~\cite{DeRujula:1975qlm},
are depicted in Fig.~\ref{meson-pse}.
The homogeneous BS equation for a pseudoscalar meson comprised of quarks with flavor $\fd$ and antiquarks with flavor $\fu$ is
\begin{eqnarray}
\nn \Gamma_{\Meps}(k;P) &=&  - \frac{16 \pi}{3} \rmh
\int \! \frac{d^4q}{(2\pi)^4} \gamma_\mu S_{\fd}(q+P) \\
&\times& \Gamma_{\Meps}(q;P)S_{\fu}(q) \gamma_\mu \,,
\label{LBSEI}
\end{eqnarray}
where $P$ is the total momentum of the bound-state. This equation has a solution for $P^2=-M_{\Meps}^2$, where $M_{\Meps}$ is the mass of the bound-state. A general decomposition for pseudoscalar mesons in the CI has the following form
 \begin{equation}
\label{KaonBSA}
\Gamma_{\Meps}(P) = i \gamma_5 \,E_{\Meps}(P) + \frac{1}{2 M_R} \gamma_5 \gamma\cdot P \, F_{\Meps}(P)\,,
\end{equation}
$M_R = M_{\fd} M_{\fu}/[M_{\fd} + M_{\fu}]$. Inserting Eq.~\eqref{KaonBSA} into Eq.~\eqref{LBSEI} and  requiring symmetry-preserving regularisation of the CI, see e.g. Ref.~\cite{Wilson:2011aa}, implies
\begin{equation}
0 = \int_0^1d\alpha \,
\left[ {\cal C}^{\rm iu}(\omega^{(1)})
+ \, {\cal C}^{\rm iu}_1(\omega^{(1)})\right], \label{avwtiP}
\end{equation}
where $\alpha$ is a Feynman parameter and
\begin{eqnarray}
\label{eq:omega}
\nn \omega^{(1)}&=&\omega(M_{\fu}^2,M_{\fd}^2,\alpha,P^2)\\
\nn &=& M_{\fu}^2 (1-\alpha) + \alpha M_{\fd}^2 + \alpha(1-\alpha) P^2\,,\\
\nn{\cal C}^{\rm iu}_1(z) &=& - z (d/dz){\cal C}^{\rm iu}(z) \\ &=& z\left[ \Gamma(0,M^2 \tau_{\rm uv}^2)-\Gamma(0,M^2 \tau_{\rm ir}^2)\right] ,\rule{2em}{0ex}
\label{eq:C1}
\end{eqnarray}
The explicit form of the BS equation is
\begin{equation}
\label{bsefinalE}
\left[
\begin{array}{c}
E_{\Meps}(P)\\
\rule{0ex}{3.0ex}
F_{\Meps}(P)
\end{array}
\right]
= \frac{4 \rmh}{3\pi}
\left[
\begin{array}{cc}
{\cal K}_{EE}^{\Meps} & {\cal K}_{EF}^{\Meps} \\
\rule{0ex}{3.0ex}
{\cal K}_{FE}^{\Meps}& {\cal K}_{FF}^{\Meps}
\end{array}\right]
\left[\begin{array}{c}
E_{\Meps}(P)\\
\rule{0ex}{3.0ex}
F_{\Meps}(P)
\end{array}
\right],
\end{equation}
with
\begin{subequations}
\label{pionKernel}
\begin{eqnarray}
\nonumber
\nn {\cal K}_{EE}^{\Meps} &=&
\int_0^1d\alpha \bigg\{
{\cal C}^{\rm iu}(\omega^{(1)})  \\
\nn&&+ \bigg[ M_{\fu} M_{\fd}-\alpha (1-\alpha) P^2 - \omega^{(1)}\bigg]
\, \overline{\cal C}^{\rm iu}_1(\omega^{(1)})\bigg\},\\
\nn {\cal K}_{EF}^{\Meps} &=& \frac{P^2}{2 M_R} \int_0^1d\alpha\, \bigg[(1-\alpha)M_{\fu}+\alpha M_{\fd}\bigg]\overline{\cal C}^{\rm iu}_1(\omega^{(1)}),\\
\nn{\cal K}_{FE}^{\Meps} &=& \frac{2 M_R^2}{P^2} {\cal K}_{EF}^K ,\\
\nn {\cal K}_{FF}^{\Meps} &=& - \frac{1}{2} \int_0^1d\alpha\, \bigg[ M_{\fu} M_{\fd}+(1-\alpha) M_{\fu}^2+\alpha M_{\fu}^2\bigg] \\
\nn &\times &\overline{\cal C}^{\rm iu}_1(\omega^{(1)})\,.
\end{eqnarray}
\end{subequations}
 \begin{table*}[ht]
\caption{\label{table-mesones-pseudo-amplitudes}
Amplitudes for pseudoscalar mesons and scalar diquark correlations (in \GeV) with the parameters listed in Table~\ref{parameters1}. $[fg]_{J^P}$ indicates a diquark formed by quarks with flavor $f$ and $g$; $J$ is the spin and $P$ the parity as usual.}
\vspace{0.0cm}
\begin{center}
\begin{tabular}{@{\extracolsep{0.85 cm}}cccccccc}
\hline
\hline
Mesons    & $\eta_c(\tc\bar{\tc})$ & $D^{0}(\tc\bar{\tu})$ & $D^{+}_{\ts}(\tc\bar{\ts})$  & $B_{\tc}^{+}(\tc\bar{\tb})$ & $B^{+}(\tu\bar{\tb})$ &  $\eta_{\tb}(\tb\bar{\tb})$ & $Bs^0(\ts\bar{\tb})$  \\
$E_{q\bar{q}}$ CI-LP &3.66 & 7.64 & 5.43& 9.55&3.10 & 4.26 & 3.77\\
$F_{q\bar{q}}$ CI-LP &1.03 & 1.08& 1.01 & 1.92&0.20 & 1.11& 0.34 \\
$E_{q\bar{q}}$ CI-HP &2.16 & 3.03 & 3.25 & 0.81 &1.50 & 0.48 & 1.59\\
$F_{q\bar{q}}$ CI-HP &0.41 & 0.25& 0.26 & 0.002 &0.007 &0.10 & 0.007 \\
\hline
Diquarks & $ [\tc\tc]_{0+}$ & $ [\tc\tu]_{0+}$& $ [\tc\ts]_{0+}$& $ [\tc\tb]_{0+}$ &  $ [\tu\tb]_{0+}$ &  $ [\tb\tb]_{0+}$  & $[\ts\tb]_{0+}$\\
$E_{qq}$ CI-LP &2.80& 4.42& 3.60 &6.58 &2.30&3.49 & 2.82\\
$F_{qq}$ CI-LP &0.73& 0.57 & 0.61 & 1.27 &0.14&0.89 & 0.25\\
$E_{qq}$ CI-HP &0.96 & 2.00 & 2.10  & 0.45 &0.98 &0.19 &0.10 \\
$F_{qq}$ CI-HP &0.19 & 0.15 & 0.16 & 0.001 &0.004 & 0.04&0.004 \\
\hline
\hline
\end{tabular}
\end{center}
\end{table*}
\noindent
Eq.~(\ref{bsefinalE}) is an eigenvalue problem. It has a solution
when $P^2=-M_{\Meps}^2$. Then the eigenvector corresponds
the BS amplitude of the meson. In the computation of observables one must employ the canonically normalized amplitude:
\begin{eqnarray}
\label{normcan}
1 &=& \left. \frac{d}{d P^2}\Pi_{\Meps}(Q,P)\right|_{Q=P}, \\
\nn\Pi_{\Meps}(Q,P)&=& 6 {\rm tr}_{\rm D} \int\! \frac{d^4q}{(2\pi)^4} \Gamma_{\Meps}(-Q)
 \frac{\partial}{\partial P_\mu} S_{\fd}(q+P)
 \\
 &\times&\Gamma_{\Meps}(Q)\, S_{\fu}(q)\,.
\end{eqnarray}
After solving the BS equation for the mesons, we can obtain the results for the diquarks.
The BS amplitude for a $J^P=0^+$-diquark made of $[\fd,\fdu]$ quarks is
\begin{equation}
\label{BSA[scalar]}
\Gamma^C_{\tdS}(P) = i \gamma_5 \,E_{\tdS}(P) + \frac{1}{2 M_R} \gamma_5 \gamma\cdot P \, F_{\tdS}(P)\,.
\end{equation}
$\Gamma^C_{\tdS}$ is the conjugate BS amplitude and satisfies
\begin{equation}
\label{bse[diquark]E}
\left[
\begin{array}{c}
E_{\tdS}(P)\\
F_{\tdS}(P)
\end{array}
\right]
= \frac{2 \rmh}{3\pi}
\left[
\begin{array}{cc}
{\cal K}_{EE}^{\Meps} & {\cal K}_{EF}^{\Meps} \\
{\cal K}_{FE}^{\Meps}& {\cal K}_{FF}^{\Meps}
\end{array}\right]
\left[\begin{array}{c}
E_{\tdS}(P)\\
F_{\tdS}(P)
\end{array}
\right].
\end{equation}
In this case the canonical normalisation condition is
\begin{equation}
1=\left. \frac{d}{d P^2}\Pi_{\tdS}(Q,P)\right|_{Q=P},
\end{equation}
where
\begin{eqnarray} \nn
\Pi_{\tdS}(Q,P)&=& 4 {\rm tr}_{\rm D} \int\! \frac{d^4q}{(2\pi)^4}\Gamma_{\tdS}(-Q)
 \frac{\partial}{\partial P_\mu} S_{\fd}(q+P) \\
& \times & \Gamma_{\tdS}(Q)\, S_{\fdu}(q)\,.
\end{eqnarray}
In Table~\ref{table-mesones-pseudo}, we present a comparison between experimental and theoretical results for the masses of pseudoscalar mesons and their diquark partners. Expectedly, CI-HP yield results closer to experimental values when
the quark masses are more disparate, as compared to the ones obtained from CI-LP. In Table~\ref{table-mesones-pseudo-amplitudes}, We list BS amplitudes for the sake of completeness and a quick
consultation.
Although the names for the mesons are the conventional ones, the quark content is also shown explicitly to avoid confusion.
An equal spacing rule  for pseudoescalar mesons containing one heavy quark is~\cite{GellMann:1962xb,Okubo:1961jc,Yin:2019bxe}
\bea \label{sumps}
m_{D_{\ts}^{+}(\tc\bar{\ts})} - m_{D^{0}(\tc\bar{\tu})} + m_{B^{+}(\tu\bar{\tb})} - m_{B_{\ts}^0(\ts\bar{\tb})} &=& 0.
\eea
The left hand side of Eq.~(\ref{sumps}) yields 0.02 GeV if we insert experimental values of the masses or the ones using CI-LP, that is, a deviation of 2\% from the sum rule, while the predictions of CI-HP yield exactly zero.
%
In heavy-light mesons, the light quark dynamics is mainly dictated by its mass and its interaction with
an almost static heavy quark. Finding a light quark away from the central core is much more probable than
the heavy one.
\subsection{Vector mesons and axial-vector diquarks}
\label{mesons-V}
\begin{figure}[H]
\vspace{-5.5 cm}
       \centerline{\includegraphics[scale=0.55,angle=0]{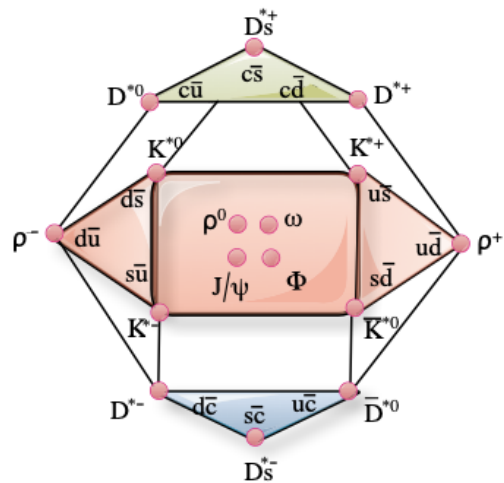}}
       \vspace{-4.8cm}
       \label{BSE}\caption{The multiplet for the vector mesons made of $\tu,\td,\ts$ and $\tc$ quarks.}
       \label{c-3d}
\end{figure}

%
In Fig.~\ref{c-3d} we depict a 15-plet and a singlet of vector mesons made of $\tu,\td,\ts$ and $\tc$ quarks. 
Following the discussion in~\cite{Roberts:2011cf}, it is straightforward to write the BS amplitude for a $J^P=1^-$- meson ($f_1 \bar{f}_2$) and $J^P=1^+$-diquark ($f_1 f_2$)~:
\begin{equation}
\label{mes-vec}
\Gamma_{\Me_{1^{-}},\mu} = \gamma_\mu^\perp E_{\Me}(P)\,,
\end{equation}
\vspace{-0.0cm}
%
\begin{table*}[htbp]
\caption{\label{table-diquarks}
Vector meson and  axial-vector diquark masses (in \GeV), computed with the parameters listed in Table~\ref{parameters1}. We present a comparison between CI and experiment~\cite{Patrignani:2016xqp,Tanabashi:2018oca}  } 
\vspace{0.0cm}
\begin{center}
\begin{tabular}{@{\extracolsep{1 cm}}cccccccc}
\hline
\hline
Mesons  & $\Jpsi$ $(\tc\bar{\tc})$ & $D^{*0}(\tc\bar{\tu})$ & $D_{\ts}^{*}(\tc\bar{\ts})$ & $B_{\tc}^{*}(\tc\bar{\tb})$  & $\Upsilon(\tb\bar{\tb})$ & $B^{+*}(\tu\bar{\tb})$ & $B_{\ts}^{0*}(\ts\bar{\tb})$ \\
Expt. &3.10 &2.01  &2.11 & $\cdots$ &9.46 &5.33 & 5.42 \\
CI-LP &2.98 &1.96& 2.05&6.18&9.46&5.11&5.20\\
CI-HP &3.14 &2.05 & 2.14 &6.31 &9.46&5.32 &5.41\\
\hline
Diquarks & $ \{\tc\tc\}_{1+}$ & $ \{\tc\tu\}_{1+}$& $ \{\tc\ts\}_{1+}$& $ \{\tc\tb\}_{1+}$& $ \{\tb\tb\}_{1+}$& $ \{\tu\tb\}_{1+}$&$ \{\ts\tb\}_{1+}$\\
CI-LP &3.12 &2.09&2.19&6.31&9.53&5.26&5.36\\
CI-HP & 3.21 & 2.16 &2.25 & 6.34&9.43 &5.38 &5.47 \\
\hline
\hline
\end{tabular}
\end{center}
\end{table*}

\begin{table*}[htbp]
\caption{\label{table-diquarks-amplitudes}
BS amplitudes required as input in FE (in \GeV) computed with the parameters listed in Table~\ref{parameters1}.} 
\vspace{0.0cm}
\begin{center}
\begin{tabular}{@{\extracolsep{0.9 cm}}cccccccc}
\hline
\hline
Mesons    & $\Jpsi$ $(\tc\bar{\tc})$ & $D^{*0}(\tc\bar{\tu})$ & $D_{\ts}^{*}(\tc\bar{\ts})$ & $B_{\tc}^{*}(\tc\bar{\tb})$  & $\Upsilon(\tb\bar{\tb})$ & $B^{+*}(\tu\bar{\tb})$ & $B_{\ts}^{0*}(\ts\bar{\tb})$ \\
$E_{q\bar{q}}$ CI-LP &3.02 & 2.39 &2.52 &3.99 &3.70 & 4.16& 4.28  \\
$E_{q\bar{q}}$ CI-HP &0.61 &1.23 & 1.31 &0.29 & 0.15 &0.65 & 0.67  \\
\hline
Diquarks & $ \{\tc\tc\}_{1+}$ & $ \{\tc\tu\}_{1+}$& $ \{\tc\ts\}_{1+}$& $ \{\tc\tb\}_{1+}$& $ \{\tb\tb\}_{1+}$& $ \{\tu\tb\}_{1+}$&$ \{\ts\tb\}_{1+}$\\
$E_{qq}$ CI-LP &2.28&1.83&1.99&3.08&3.02&2.97&3.05\\
$E_{qq}$ CI-HP &0.41 &0.92 &0.95 &0.21 &0.09 &0.48&0.48 \\
\hline
\hline
\end{tabular}
\end{center}
\end{table*}
The explicit form of Eq.~(\ref{genbse}) for the ground-state vector meson, whose solution yields its mass-squared, is
where $P_\mu \gamma_\mu^\perp = 0$. The explicit form of Eq.~(\ref{genbse}) for the ground-state vector-meson, whose solution yields its mass-squared, is
\begin{equation}\label{meson-bse}
1 - {\cal K}^{\Me}(-m_{\Me}^2) = 0\,,
\end{equation}
with
\begin{eqnarray}\nn
\label{KastKernel} {\cal K}^{\Me}(P^2)= \frac{2\rmh}{3\pi} \int_0^1d\alpha\,
{\cal L}^{\Me}(P^2)
\overline{\cal C}_1^{\rm iu}(\omega(M_{\fu}^2, M_{\fd}^2, \alpha, P^2))
\end{eqnarray}
and
\bea \nn {\cal L}^{\Me}(P^2)= M_{\fu} M_{\fd} - (1-\alpha)M_{\fu}^2-\alpha M_{\fd}^2-2\alpha(1-\alpha)P^2 . \eea
The functions $\omega(M_{\fu}^2,M_{\fd}^2,\alpha,P^2)$ and ${\cal C}^{\rm iu}_1(z) $ are
\begin{eqnarray}
\label{eq:omega}
\nn&& \hspace{-0.4cm} \omega(M_{\fu}^2,M_{\fd}^2,\alpha,P^2) =M_{\fu}^2 (1-\alpha) + \alpha M_{\fd}^2 + \alpha(1-\alpha) P^2\,,\\
&& \hspace{-0.4cm} \nn{\cal C}^{\rm iu}_1(z) = - z (d/dz){\cal C}^{\rm iu}(z) = z\left[ \Gamma(0,M^2 \tau_{\rm uv}^2)-\Gamma(0,M^2 \tau_{\rm ir}^2)\right] .\rule{2em}{0ex}
\label{eq:C1}
\end{eqnarray}
The corresponding canonical normalisation condition can be written as~:
\begin{equation}
\frac{1}{E_{\Me}^2} = 9 {m}_G^2 \left. \frac{d}{d z} {\cal K}^{\Me}(z)\right|_{z=-m_{\Me}^2} \,.
\end{equation}
%
%
Following this pattern one may immediately write down the BS equation for $J^P=1^+$ diquark correlations~:
\begin{equation}
1 - \frac{1}{2} K^{K^{\tavd}}(-m_{\tavd_{1^+}}^2) = 0\,.
\end{equation}
We adopt the notation $\tavd$ for axial-vector diquarks. The canonical normalisation condition is~:
\begin{equation}
\label{canonicalavqq}
\frac{1}{E_{\tavd_{1^+}}^2} = 6 {m}_G^2 \left.\frac{d}{d z} K^{\tavd}(z)\right|_{z=-m_{\tavd_{1^+}}}.
\end{equation}
The results are reported in Tables~\ref{table-diquarks} and~\ref{table-diquarks-amplitudes}.

Equal spacing rule for  vector mesons with the same quark content in Eq.~(\ref{sumps}),~\cite{GellMann:1962xb,Okubo:1961jc,Yin:2019bxe}
\bea \label{sumv}
m_{D_{\ts}^{*}(\tc\bar{\ts})} - m_{D^{*0}(\tc\bar{\tu})} + m_{B^{+*}(\tu\bar{\tb})} - m_{B_{\ts}^{0*}(\ts\bar{\tb})} = 0.
\eea
The Eq.~(\ref{sumv}) is exactly fulfilled for the experimental results and for the HP. The results using LP have an error of 1\%. In the same references~\cite{GellMann:1962xb,Okubo:1961jc,Yin:2019bxe}, we can also find
approximate mass relations connecting vector and pseudoscalar mesons~:
\bea \label{gmo-1}
&&\hspace{-0.7cm}m_{B_c^{*}(\tc\bar{\tb})}-m_{B_s^{0*}(\ts\bar{\tb})}-m_{B_c^{+}(\tc\bar{\tb})}+m_{B_s^{0}(\ts\bar{\tb})}\approx 0\,,\\
\label{gmo-2}&& \hspace{-0.7cm}m_{B_{\ts}^{0*}(\ts\bar{\tb})}-m_{B^{+*}(\tu\bar{\tb})}-m_{B_{\ts}^0(\ts\bar{\tb})}+m_{B^{+}(\tu\bar{\tb})}= 0\,,\\
\label{gmo-3}&&\hspace{-0.7cm}m_{B_{\ts}^{0*}(\ts\bar{\tb})}-m_{B^{+*}(\tu\bar{\tb})}-m_{D_{\ts}^{+}(\tc\bar{\ts})}+m_{D^{0}(\tc\bar{\tu})}=0\,, \\
\label{gmo-4}&&\hspace{-0.7cm}m_{\eta_{\tb}(\tb\bar{\tb})}-m_{\eta_{\tc}(\tc\bar{\tc})}-2m_{B_{\ts}^{0*}(\ts\bar{\tb})}+2m_{D_{\ts}^{*}(\tc\bar{\ts})}\approx 0\,,\\
\label{gmo-5}&&\hspace{-0.7cm}m_{\eta_{\tb}(\tb\bar{\tb})}-m_{\eta_{\tc}(\tc\bar{\tc})}-2m_{B_{\ts}^0(\ts\bar{\tb})}+2m_{D_{\ts}^{+}(\tc\bar{\ts})}=0\,,\\
\label{gmo-6}&&\hspace{-0.7cm}m_{B_{\ts}^{0*}(\ts\bar{\tb})}-m_{D_{\ts}^{*}(\tc\bar{\ts})}-m_{B_{\ts}^0(\ts\bar{\tb})}+m_{D_{\ts}^{+}(\tc\bar{\ts})}=0\,.\\
\label{gmo-7}&&\hspace{-0.7cm}m_{\Upsilon(\tb\bar{\tb}) }-m_{\Jpsi(\tc\bar{\tc})}-2m_{B_{\ts}^0(\ts\bar{\tb})}+2m_{D_{\ts}^{+}(\tc\bar{\ts})}=0\,.\\
\label{gmo-8}&&\hspace{-0.7cm}m_{\Upsilon(\tb\bar{\tb}) }-m_{\Jpsi(\tc\bar{\tc})}-m_{\eta_{\tb}(\tb\bar{\tb})}+m_{\eta_{\tc}(\tc\bar{\tc})}\approx 0\,.\\
\label{gmo-9}&&\hspace{-0.7cm}m_{\Upsilon(\tb\bar{\tb}) }-m_{\Jpsi(\tc\bar{\tc})}-2m_{B_{\ts}^{0*}(\ts\bar{\tb})}+2m_{D_{\ts}^{*}(\tc\bar{\ts})}\approx 0\,.
\eea
We test these mass relations, Eqs.~(\ref{gmo-1}-\ref{gmo-9}), against experiment. The deviation from these mass relations is listed in the table below:
\vspace{-1.0cm}
\begin{center}
\begin{tabular}{@{\extracolsep{0.4 cm}}cccc}
\hline
\hline
Spacing rule & CI-LP & CI-HP & Expt. \\
\hline
\rule{0ex}{3.0ex}
 Eq.~(\ref{gmo-1}) &-0.04 & -0.01 &  $\cdots$\\
 \rule{0ex}{3.0ex}
 Eq.~(\ref{gmo-2}) &-0.01 & 0 & 0\\
 \rule{0ex}{3.0ex}
 Eq.~(\ref{gmo-3}) &-0.03 & 0 &-0.02 \\
 \rule{0ex}{3.0ex}
 Eq.~(\ref{gmo-4}) &0.18 & -0.07 &-0.2\\
 \rule{0ex}{3.0ex}
 Eq.~(\ref{gmo-5}) &0.02 &-0.37 &-0.38 \\
 \rule{0ex}{3.0ex}
 Eq.~(\ref{gmo-6}) &-0.08 &-0.15 &-0.09 \\
 \rule{0ex}{3.0ex}
 Eq.~(\ref{gmo-7}) &0.02 &-0.52 &-0.44 \\
 \rule{0ex}{3.0ex}
 Eq.~(\ref{gmo-8}) &0 &-0.15 &-0.06 \\
 \rule{0ex}{3.0ex}
 Eq.~(\ref{gmo-9}) &0.18 &-0.22 &-0.26 \\
 \hline
 \hline
\end{tabular}
\end{center}
%
\section{Heavy Baryons}
\label{Baryons}

\begin{figure}[ht]
       \vspace{-5.0cm}
\centerline{\includegraphics[scale=0.6,angle=0]{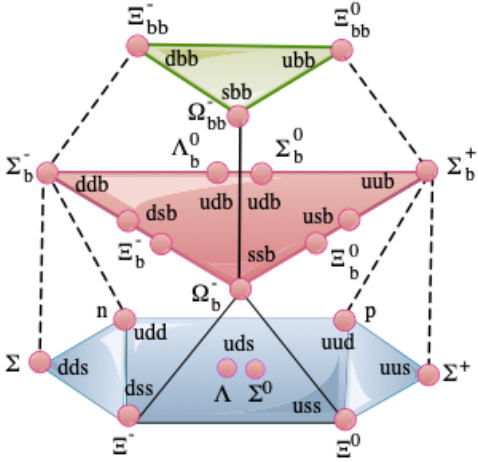}}
       \vspace{-5.3cm}
       \caption{We show the mixed-symmetric $20$-plet. Note that all the ground-state baryons in this multiplet have $J^P =(1/2)^+$. It has the SU(3) octet on the lowest layer. The singly heavy baryons are composed of a bottom quark and two light quarks $(\tu,\td,\ts)$, located in the second layer. The doubly heavy baryons are positioned in the top-most layer}.
       \label{Bary-bb}
       \end{figure}

\begin{figure}[htbp]
\vspace{-4.5 cm}
      \centerline{ \includegraphics[scale=0.6,angle=0]{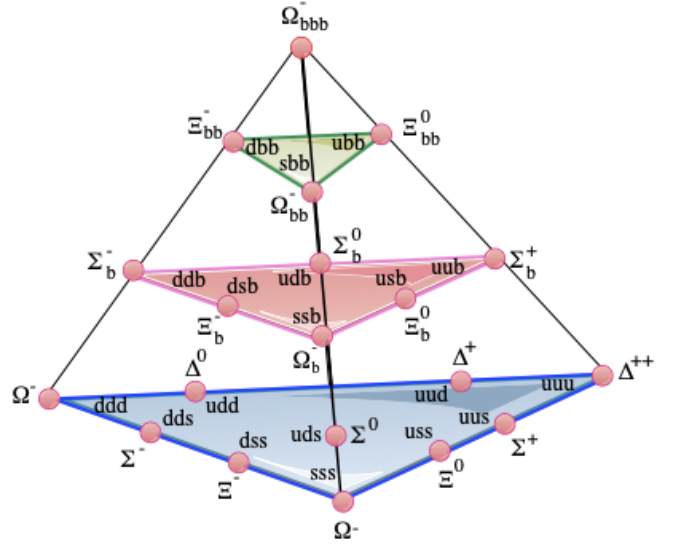}}
       \vspace{-4.8cm}
       \caption{ States of baryons with spin $3/2$ made from four quarks of the types $\tu, \td, \ts$, and $\tb$. Doubly heavy baryons and triply heavy baryons are localized in the highest layers}
       \label{Bary-b}
\end{figure}

\begin{figure}[htpb]
\vspace{-3.3cm}
       \centerline{\includegraphics[scale=0.42,angle=-90]{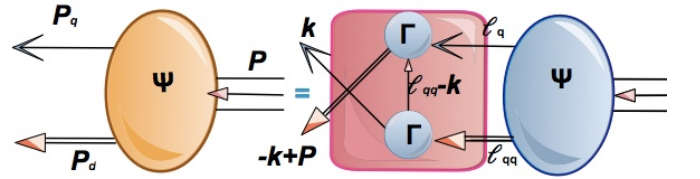}}
       \vspace{-2.8cm}\caption{Poincar\'e covariant FE employed in this work to calculate baryon masses. The square represents the quark-diquark interaction Kernel. The single line denotes the dressed quark propagator, the double line is the diquark propagator while $\Gamma$ and $\Psi$ are the BS and Faddeev amplitudes, respectively. Configuration of momenta is: $\ell_{qq}=-\ell + P$, $k_{qq}=-k+P$, $P=P_d+P_q$}.
       \label{faddevv-Fig1}
\end{figure}

The $SU(5)$ flavor group includes all types of baryons containing zero, one, two or three heavy quarks.
The baryons multiplets that arise from $3\otimes3\otimes3$ are: a decuplet, two octets and a singlet.
The corresponding multiplet structure for $SU(4)$ is $4\otimes4\otimes 4 = 20_S\oplus20_M\oplus20_M\oplus 4_A$.
Note that explicit quark masses break the flavor symmetry. The larger the group, the bigger is the amount
of breaking. However, the group algebra helps us identify the baryons whose masses we will compute.
As an example, we present such baryon multiplets with $\tu$, $\td$, $\ts$ and $\tb$ quark in Figs.~\ref{Bary-bb}, and~\ref{Bary-b}. The multiplet with charm quarks is analogous to the one containing the bottom quark.
\vspace{-0cm}
\vspace{-0 cm}
 \\
Here, we consider the three-quark systems with one, two or
three heavy quarks. The singly heavy baryons are not all discovered. A lot of
literature is available on theoretical studies concerning the doubly and even triply heavy baryons with different approaches, including Poincaré-covariant analysis(PC) of continuum QCD~\cite{Qin:2018dqp},~lattice~\cite{Brown:2014ena,Mathur:2018epb},~variational Coulomb and Cornell potentials~\cite{Jia:2006gw}, Faddeev equation formalism (Fadv)~\cite{SilvestreBrac:1996bg}, the bag model (BM)~\cite{Hasenfratz:1980ka}, quark countig rules (QCR)~\cite{Bjorken:1985ei}, constituent quark model (CQM1)~\cite{Roberts:2007ni} and (CQM2)~\cite{Vijande:2004at}, relativistic quark model (RQM)~\cite{Martynenko:2007je}, instanton quark model (IQM)~\cite{Migura:2006ep}, hypercentral model (HCM)~\cite{Patel:2008mv}, QCD sum rules (SR)~\cite{Zhang:2009re}, Regee phenomenology~\cite{Guo:2008he} and non-relativistic QCD (NRQCD)~\cite{LlanesEstrada:2011kc}.
On the experimental side, in 2002, the SELEX Collaboration~\cite{Mattson:2002vu} reported the
first observation of a doubly charmed baryon $\Xi_{cc}^{+}$ in the
decay mode $\Xi_{cc}^{+}\to\Lambda_c^{+}K^{-}\pi^{+}$.
Its mass was determined to be $3519 \pm 1\,\rm MeV$.
Further works identified its isospin partner $\Xi_{cc}^{++}(3460)$~\cite{Russ:2002bw}.
Recently, the LHCb Collaboration~\cite{Aaij:2017ueg} reported the
observation of $\Xi_{cc}^{++}$ in the $\Lambda_c^+K^-\pi^+\pi^+$ decay
mode.
But its mass was determined to be $ 3621.40 \pm 0.72 (\text{stat.})
\pm 0.27 (\text{syst.}) \pm 0.14 (\Lambda_c^+)\,\rm MeV $. With the ongoing theoretical efforts and experimental discoveries, we join the timely effort to study these
baryons.
%
%
%
%
In this section we extend the CI model in the heavy baryon sector. We compute the masses of positive parity spin-$1/2$ and $3/2$ baryons composed of $\tu$, $\td$, $\ts$, $\tc$ and $\tb$ quarks in a quark-diaquark picture employing both the CI-LP and CI-HP parameters.
We base our description of baryon bound-states on FE, which
is illustrated in Fig.~\ref{faddevv-Fig1}.
\vspace{0 cm}
\vspace{-1.5cm}
\subsection{Baryons with spin $1/2$}
\label{baryons-1}
The mass of the ground-state baryon with spin $1/2$ comprised by the quarks $[\tq\tq\tqu]$ is  determined by a $5\times 5$ matrix FE.
One can write it in the following form~:
\begin{eqnarray}
\nonumber
\lefteqn{
 \left[ \begin{array}{r}
{\cal S}(k;P)\, u(P)\\
{\cal A}^i_\mu(k;P)\, u(P)
\end{array}\right]}\\
& =&  -\,4\,\int\frac{d^4\ell}{(2\pi)^4}\,{\cal M}(k,\ell;P)
\left[
\begin{array}{r}
{\cal S}(\ell;P)\, u(P)\\
{\cal A}^j_\nu(\ell;P)\, u(P)
\end{array}\right] .\rule{1em}{0ex}
\label{FEone}
\end{eqnarray}
The general matrices ${\cal S}(\ell;P)$ and ${\cal A}^i_\nu(\ell;P)$, which describe the momentum-space correlation between the quark and diquark in the nucleon and the Roper, are described in Refs.~\cite{Oettel:1998bk, Cloet:2007pi}.  However, with the interaction employed in this article, they simplify considerably
\begin{subequations}
\label{FaddeevAmp}
\begin{eqnarray}
{\cal S}(P) &=& s(P) \,\mbox{\boldmath $I$}_{\rm D}\,,\\
{\cal A}^i_\mu(P) &=& a_1^i(P) \gamma_5\gamma_\mu + a_2^i(P) \gamma_5 \hat P_\mu \,, \quad i=+,0\rule{2em}{0ex}
\end{eqnarray}
\end{subequations}
where the scalars $s$ and $a_{1,2}^i$ are independent of the relative quark-diquark momentum and $\hat P^2=-1$. The Faddeev amplitude is thus represented by the eigenvector~:
\begin{equation}
\Psi(P) = \left[\begin{array}{c}
s(P)\\[0.7ex]
a_1^+(P)\\[0.7ex]
a_1^0(P)\\[0.7ex]
a_2^+(P)\\[0.7ex]
a_2^0(P)\end{array}\right] \,.
\end{equation}
The kernel in Eq.~(\ref{FEone}) is
\begin{eqnarray}
\label{calM} {\cal M}(k,\ell;P) =
 \begin{pmatrix} \mathcal{M}^{11}&\mathcal{M}^{12}_\nu&\mathcal{M}^{13}_\nu&\mathcal{M}^{14}_\nu&\mathcal{M}^{15}_\nu\\ \\
 \mathcal{M}^{21}_\mu& \mathcal{M}^{22}_{\mu\nu}&\mathcal{M}^{23}_{\mu\nu}& \mathcal{M}^{24}_{\mu\nu}&\mathcal{M}^{25}_{\mu\nu}\\ \\
 \mathcal{M}^{31}_{\mu}&\mathcal{M}^{32}_{\mu\nu}& \mathcal{M}^{33}_{\mu\nu}&\mathcal{M}^{34}_{\mu\nu}&\mathcal{M}^{35}_{\mu\nu}\\ \\
 \mathcal{M}^{41}_\mu& \mathcal{M}^{42}_{\mu\nu}&\mathcal{M}^{43}_{\mu\nu}&\mathcal{M}^{44}_{\mu\nu}&\mathcal{M}^{45}_{\mu\nu}\\ \\
 \mathcal{M}^{51}_{\mu}&\mathcal{M}^{52}_{\mu\nu}& \mathcal{M}^{53}_{\mu\nu}&\mathcal{M}^{54}_{\mu\nu}&\mathcal{M}^{55}_{\mu\nu} \\
 \label{B-matrix} \end{pmatrix} \,. \end{eqnarray}
The elements of the matrix in~(\ref{B-matrix}) are detailed in appendix~\ref{app:Fad}. In order to simplify Eqs.~(\ref{FEone}), we use static approximation for the exchanged quark with flavor $f$. It was introduced long ago in Ref.~\cite{Buck:1992wz}
\bea
S(p)=\frac{1}{i \gamma \cdot p + M_f} \to
\frac{1}{M_f} \,. \eea
A variation of it was implemented in~\cite{Xu:2015kta},
\bea
S(p)=\frac{1}{i \gamma \cdot p + M_f}\to
\frac{g^2_{N\,\Delta}}{i \gamma \cdot p + M_f}  \,. \eea
We follow refs.~\cite{Roberts:2011cf,Lu:2017cln,Chen:2012qr} and represent the quark (propagator) exchanged between the diquarks as
\begin{equation}
S^{\rm T}(k) \to \frac{g_B}{M_f}\,.
\label{staticexchangec}
\end{equation}
The superscript ``T'' indicates matrix transpose.
In the implementation of this treatment for heavy baryons with spin-$1/2$ we use $g_B=0.75$ for CI-LP and $g_B=1$ for CI-HP. Explicit expressions for the flavor matrices $t$ for the diquark pieces can be found in Appendix~\ref{app:Fla}. The spin-$1/2$ heavy baryons are represented by the
following column matrices:
 \begin{equation}
\nn \begin{array}{cc}
u_{\Xi_{cc}^{++} (\tu\tc\tc)}= \left[
\begin{array}{c}
[\tu\tc] \tc \\
\{\tc\tc\}\tu\\
\{\tu\tc\}\tc\\
\end{array} \right], &
u_{\Omega_{\tc\tc}^+(\ts\tc\tc)}=
\left[ \begin{array}{c}
[\ts\tc] \tc \\
\{\tc\tc\} \ts \\
\{\ts\tc\} \tc \\
\end{array} \right],
\end{array}
\end{equation}

 \begin{equation}
\nn \begin{array}{cc}
u_{\Omega_{\tc}^0(\ts\ts\tc)}= \left[
\begin{array}{c}
[\ts\tc] \ts \\
\{\ts\ts\}\tc\\
\{\ts\tc\}\ts\\
\end{array} \right], &
u_{\Sigma_{\tc}^{++}(\tu\tu\tc)}=
\left[ \begin{array}{c}
[\tu\tc] \tu \\
\{\tu\tu\} \tc \\
\{\tu\tc\} \tu \\
\end{array} \right],
\end{array}
\end{equation}

 \begin{equation}
\nn \begin{array}{cc}
u_{\Xi_{bb}^0(\tu\tb\tb)}= \left[
\begin{array}{c}
[\tu\tb] \tb \\
\{\tb\tb\}\tu\\
\{\tu\tb\}\tb\\
\end{array} \right], &
u_{\Omega_{bb}^-(\ts\tb\tb)}=
\left[ \begin{array}{c}
[\ts\tb] \tb \\
\{\tb\tb\} \ts \\
\{\ts\tb\} \tb \\
\end{array} \right],
\end{array}
\end{equation}
 \begin{equation}
\nn \begin{array}{cc}
u_{\Omega_b^{-}(\ts\ts\tb)}= \left[
\begin{array}{c}
[\ts\tb] \ts \\
\{\ts\ts\}\tb\\
\{\ts\tb\}\ts\\
\end{array} \right], &
u_{\Sigma_b^+(\tu\tu\tb)}=
\left[ \begin{array}{c}
[\tu\tb] \tu \\
\{\tu\tu\} \tb \\
\{\tu\tb\} \tu \\
\end{array} \right],
\end{array}
\end{equation}
 \begin{equation}
\nn \begin{array}{cc}
u_{\Omega(\tc\tc\tb)}= \left[
\begin{array}{c}
[\tc\tb] \tc \\
\{\tc\tc\}\tb\\
\{\tc\tb\}\tc\\
\end{array} \right], &
u_{\Omega(\tc\tb\tb)}=
\left[ \begin{array}{c}
[\tc\tb] \tb \\
\{\tb\tb\} \tc \\
\{\tc\tb\} \tb \\
\end{array} \right],
\end{array}
\end{equation}
Experimental and calculated masses of spin $1/2$-baryons with
charm and bottom quarks are listed in
Table~\ref{table-baryons-oct}, specifying the percentage difference between
them.
One of the most striking results is that for the mass of the $\Xi_{\tc\tc}^{++}$ baryon observed in LHCb. The difference the our work and the reported experimental value is $0.055\%$ with CI-HP and zero with CI-LP. This result is also consistent with the spacing rule for heavy baryons with one heavy and two light quarks~\cite{Ebert:2005xj,GellMann:1962xb,Okubo:1961jc}
\bea
&&m_{\Sigma_Q}+m_{\Omega_Q}=2m_{\Xi_Q} \;. \;\;\;\;\;\;\;\;\;Q=\tc,\;\tb
\eea
We test this rule for the following baryons~:
\begin{eqnarray}
\label{gm-l1} m_{\Sigma_{\tc}^{++}(\tu\tu\tc)}+m_{\Omega_{\tc}^0(\ts\ts\tc)} &=& 2m_{\Xi_{\tc}^+{(\tu\ts\tc)}}\\
\label{gm-l2} m_{\Sigma_{\tb}^+{(\tu\tu\tb)}}+m_{\Omega_{\tb}^-{(\ts\ts\tb)}} &=& 2m_{\Xi_{\tb}^0{(\tu\ts\tb)}} \;.
\end{eqnarray}
Using Eq.~(\ref{gm-l1}) with CI-LP and CI-HP, we find $m_{\Xi_{\tc}^+{(\tu\ts\tc)}}=2.70$ $\GeV$. The experimental value is $2.47$ $\GeV$~\cite{Tanabashi:2018oca}.
On the other hand, the experimental value for $m_{\Xi_{\tb}^0{(\tu\ts\tb)}}$ is $5.79$ $\GeV$~\cite{Tanabashi:2018oca} while the CI model yields $5.91$ and $5.90$ $\GeV$ for CI-LP and CI-HP, respectively.
Both sets of parameters are reasonably consistent with the measurements.
For the sake of completeness, we present the corresponding Faddeev amplitudes in Table~\ref{fbo}.
 \begin{table*}[ht]
\caption{\label{table-baryons-oct} Baryons with spin 1/2. First row shows the values with lattice QCD and the last two rows with the label Diff. indicates the percentage of difference between the experimental results and those computed with CI~\cite{Brown:2014ena,Mathur:2018epb}  } 
\vspace{0 cm}
\begin{tabular}{@{\extracolsep{0.3 cm}}ccccccccccc}
\hline
\hline
&$\Xi_{cc}^{++} (\tu\tc\tc)$ & $\Omega_{\tc\tc}^+(\ts\tc\tc)$ & $\Omega_{\tc}^0(\ts\ts\tc)$ & $\Sigma_{\tc}^{++}(\tu\tu\tc)$& $\Xi_{bb}^0(\tu\tb\tb)$ & $\Omega_{bb}^-(\ts\tb\tb)$ & $\Omega_b^{-}(\ts\ts\tb)$ & $\Sigma_b^+(\tu\tu\tb)$& $\Omega(\tc\tb\tb)$ & $\Omega(\tc\tc\tb)$ \\
Expt. & 3.62 & $\cdots$ & 2.69 & 2.45 & $\cdots$& $\cdots$ & 6.04&5.81& $\cdots$ &  $\cdots$  \\
Lattice &3.61 & 3.74 &2.68 &2.43 &10.14 &10.27 &6.06 &5.86 &11.19 & 8.01
\\
CI-LP & 3.62 & 3.80& 2.85 & 2.55&10.08&10.19&6.03&5.79&11.16&7.97 \\
CI-HP & 3.64 & 3.76 & 2.82 & 2.58 & 10.06 & 10.14 & 6.01 & 5.78 & 11.09 & 8.01 \\
\hline
Diff. CI-LP &0\%& $\cdots$ & 5.94\%&4.08\% &  $\cdots$ & $\cdots$ &0.16\% & 0.34\%&  $\cdots$ &  $\cdots$\\
Diff. CI-HP &0.55\%& $\cdots$  & 4.83\%&5.30\% & $\cdots$ & $\cdots$ &0.49\% & 0.51\%& $\cdots$ & $\cdots$\\
\hline
\hline
\end{tabular}
\end{table*}
\begin{table*}[htp]
\caption{\label{fbo} CI Faddeev amplitudes. It is is notable that the amplitudes give the dominant diquark within the baryon. In the last column we indicate this diquark in each case.} 
\vspace{0 cm}
\begin{tabular}{@{\extracolsep{0.4 cm}}ccccccc|c}
\hline
\hline
& & $s$ & $a_1^+$ & $a_1^0$ & $a_2^+$ & $a_2^0$ & dom \\
 \rule{0ex}{3.0ex}
 \multirow{2}{2.0cm}{$\Xi_{cc}^{++} (\tu\tc\tc)$ } &CI-LP &0.81 &-0.15 & 0.35& -0.35& -0.29 & $[\tu\tc]\tc$
\\ \rule{0ex}{3.0ex}
& CI-HP  & -0.88 & 0.05 &
-0.36&
0.11 &
0.30 & $[\tu\tc]\tc$ \\
%
\hline
\rule{0ex}{3.0ex}
\multirow{2}{2.0cm}{$\Omega_{\tc\tc}^+(\ts\tc\tc)$  } &CI-LP & 0.66 & -0.20 & 0.34 & -0.55 & -0.31 & $[\ts\tc]\tc$
\\ \rule{0ex}{3.0ex}
& CI-HP  & -0.88 &
0.07 &
-0.35 &
0.11 &
0.30 & $[\ts\tc]\tc$
\\
\hline
\rule{0ex}{3.0ex}
\multirow{2}{2.0cm}{$\Omega_{\tc}^0(\ts\ts\tc)$} &CI-LP &  -0.34 & 0.26 &-0.12 & -0.90 &0.01 & $\{\ts\ts\}\tc$
\\
\rule{0ex}{3.0ex}
& CI-HP  & 0.57 &
-0.29 &
0.10 &
0.76 &
-0.02 &
$\{\ts\ts\}\tc$
\\
\hline
%
\rule{0ex}{3.0ex}
\multirow{2}{2.0cm}{$\Sigma_{\tc}^{++}(\tu\tu\tc)$ } &CI-LP  & -0.35 & 0.24 & -0.09 &-0.90 & 0.01 & $\{\tu\tu\}\tc$
\\ \rule{0ex}{3.0ex}
& CI-HP & -0.49&
0.26 &
-0.09 &
-0.82 &
0.01 &
$\{\tu\tu\}\tc$
\\
\hline
\rule{0ex}{3.0ex}
\multirow{2}{2.0cm}{$\Xi_{bb}^0(\tu\tb\tb)$ } &CI-LP & 0.18 & -0.02 & 0.12 & -0.97& -0.11 & $\{\tb\tb\}\tu$
\\ \rule{0ex}{3.0ex}
& CI-HP & -0.11 &
0.07 &
-0.04 &
0.99 &
0.06 &
$\{\tb\tb\}\tu$
\\
\hline
\rule{0ex}{3.0ex}
\multirow{2}{2.0cm}{ $\Omega_{bb}^-(\ts\tb\tb)$} &CI-LP & -0.16 & 0.02 & -0.09 & 0.98 & 0.08& $\{\tb\tb\}\ts$
\\ \rule{0ex}{3.0ex}
& CI-HP & 0.12 &
-0.10 &
0.04 &
-0.98 &
-0.07 &
$\{\tb\tb\}\ts$
\\
\hline
%
 \rule{0ex}{3.0ex}
 \multirow{2}{2.0cm}{  $\Omega_b^{-}(\ts\ts\tb)$} &CI-LP & -0.08 & 0.13 & -0.05 & -0.99 & -0.03 & $\{\ts\ts\}\tb$
\\ \rule{0ex}{3.0ex}
& CI-HP & 0.12 &
-0.10 &
0.04 &
-0.98 &
-0.07 &
$\{\ts\ts\}\tb$
\\
\hline
 \rule{0ex}{3.0ex}
  \multirow{2}{2.0cm}{  $\Sigma_b^+(\tu\tu\tb)$ } & CI-LP & -0.06 & 0.11 & -0.05 & -0.99 & -0.03 & $\{\tu\tu\}\tb$ 
\\ \rule{0ex}{3.0ex}
& CI-HP& 0.5 &
-0.20&
0.05 &
0.83 &
0.04 &
$\{\tu\tu\}\tb$
\\
\hline
 \rule{0ex}{3.0ex}
  \multirow{2}{2.0cm}{ $\Omega(\tc\tb\tb)$ } & CI-LP & 0.15 &
-0.03&
0.06&
-0.98&
-0.05&
$\{\tb\tb\}\tc$ 
\\ \rule{0ex}{3.0ex}
& CI-HP& -0.77&
0.05&
-0.30 &
0.49&
0.28&
$[\tc\tb]\tb$
\\
\hline
 \rule{0ex}{3.0ex}
  \multirow{2}{2.0cm}{ $\Omega(\tc\tc\tb)$ } & CI-LP &-0.38 &
0.23 &
-0.09 &
-0.89&
-0.02&
$\{\tc\tc\}\tb$ 
\\ \rule{0ex}{3.0ex}
& CI-HP& -0.82 &
0.21&
-0.008&
-0.53&
-0.003&
$\{\tc\tc\}\tb$
\\
\hline
\hline
\end{tabular}
\end{table*}
 %
\subsection{baryons with spin $3/2$}
\label{baryons-2}

\begin{figure}[H]
\vspace{-0.8cm}
\centerline{
       \includegraphics[scale=.35,angle=-90]{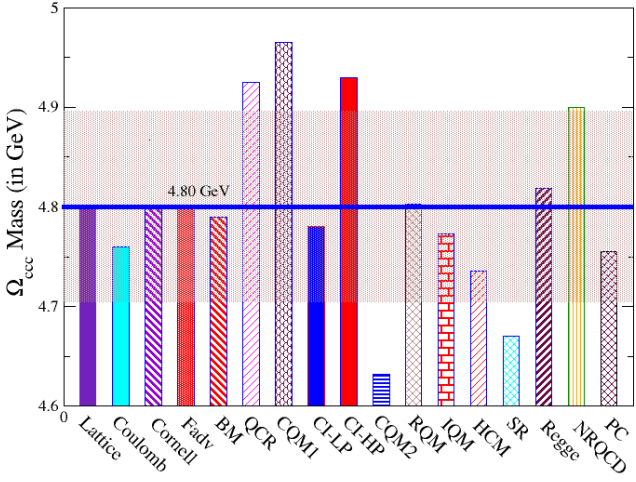}}
       \caption{\label{omegac1} We plot the mass of $\Omega_{\tc\tc\tc}$ using CI and other approaches. The blue line indicates the average. The shaded section shows 2\% difference with the average.}
       \label{BSEfig}
\end{figure}

\begin{figure}[htpb]
\vspace{-1.0cm}
\centerline{
       \includegraphics[scale=.350,angle=-90]{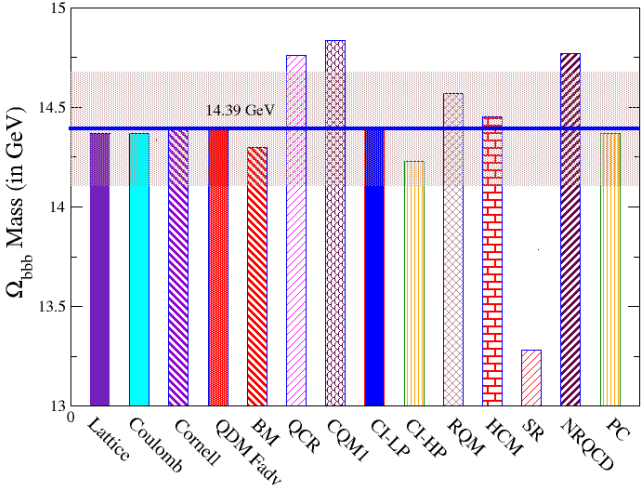}}
       \caption{\label{omegab1} Our prediction for the $\Omega_{\tb\tb\tb}$.  We  compare our result with a number
       of other computations in literature.}
       \label{BSEfig}
\end{figure}

Baryons with spin $3/2$ are especially important because they can involve states with three $\tc$-quarks and three $\tb$-quarks. In order to calculate the masses we note that it is not possible to combine a spin-zero
diquark with a spin-1/2 quark to obtain spin-$3/2$ baryon. Hence such a baryon is comprised solely of axial-vector correlations.
The Faddeev amplitude for the positive-energy baryon
is~:
\bea
\Psi_\mu = \psi_{\mu\nu}(P)u_\nu(P) \;, \nn
\eea
where $P$ is the baryon's total momentum and $u_\nu(P)$ is a Rarita-Schwinger spinor,
\bea \label{fd1}
\psi_{\mu\nu}(P)u_{\nu} &=& \Gamma_{qq_{1^+} \mu} \Delta_{\mu\nu,_{qq}}^{1^+}(\ell_{qq}){\cal D}_{\nu\rho}(P)u_{\rho}(P) \,.
\eea
Understanding the structure of these states is simpler than in the case of the nucleon and
\bea
\label{DeltaFA}
{\cal D}_{\nu\rho}(\ell;P) &=& {\cal S}(\ell;P) \, \delta_{\nu\rho} + \gamma_5{\cal A}_\nu(\ell;P) \,\ell^\perp_\rho \,.
\eea
We give more details of this equation in the  appendix~\ref{App:EM}.
%
We will consider the baryons with two possible structures: $\tq\tq\tq$ and $\tqu\tq\tq$. \\ \\
%
\noindent
{\bf Baryons($\tq\tq\tq$):} A single possible combination of diquarks exists for a baryon composed of the same three quarks $(\tq\tq\tq)$. In this case, the Faddeev amplitude is:
\begin{equation}
{\cal D}_{\nu\rho}(\ell;P) u_\rho^B(P) = f^B(P) \, \mathbf{I}_{\rm D} \, u_\nu^B(P)\,.
\label{DnurhoI}
\end{equation}
Employing Feynman rules for Fig.~\ref{faddevv-Fig1} and using the expression for the Faddevv amplitude, eq.~({\ref{DnurhoI}}), we can write
\bea
f^B(P) u_\mu^B(P)=
4\frac{g_B}{M_{\tq}}\int \frac{d^4 \ell}{(2\pi)^4}\mathcal{M}f^B(P) u_\nu^B(P) \,,
\eea
where we have suppressed the functional dependence of $\mathcal{M}$ on
momenta for the simplicity of notation. We
now multiply both sides by ${\bar{u}}^B_{\beta}(P)$ from the left and sum over the polarization
not explicitly shown here, to obtain
\bea
\Lambda_{+}(P)R_{\mu\beta}(P)=4\frac{g_B}{M_{\tq}}\int \frac{d^4 \ell}{(2\pi)^4}\mathcal{M}\Lambda_{+}(P)R_{\nu\beta}
\eea
Finally we contract with $\delta_{\mu\beta}$
\begin{eqnarray} \nn  \label{o-diquark}
2\pi^2 = \frac{1}{M_{\tq}}\frac{E_{\{\tq\tq\}_{1^+}}^2}{m_{\{\tq\tq\}_{1^+}}^2}
 \hspace{-2mm} \int_0^1 \hspace{-2mm} d\alpha\, {\cal L}^\Omega \overline{\cal C}^{\rm iu}_1(\omega (\alpha,M_{\tq}^2,m^2_{\{\tq\tq\}_{1^+}},m_B^2))\,,
\end{eqnarray}
where we have defined
\bea
{\cal L}^\Omega= [m_{\{\tq\tq\}_{1^+}}^2 + (1-\alpha)^2 m_\Delta^2][\alpha m_\Delta + M_{\tq}]
\,.
\eea
From the last two expressions, it is straightforward to compute the mass of the baryon constituted
by three equally heavy quarks. \\ \\
\noindent
{\bf Baryons($\tqu\tq\tq$):}
 For a baryon with quark structure ($\tqu\tq\tq$), there are two possible diquarks, $\{\tq\tq\}$ and $\{\tqu\tq\}$. The Faddeev amplitude for such a baryon is:
\begin{equation}  \label{d-diquark}
{\cal D}_{\nu\mu}^{B}(P)u_{\mu}^{B}(P;s) =
\sum_{i=\{\tqu\tq\},\{\tq\tq\}} d^{i}(P) \delta_{\nu\lambda} u_{\lambda}^{B}(P;s),
\end{equation}
so that the corresponding FE has the form
\begin{equation}
\left[\begin{matrix}
d^{\{\tqu\tq\}}\\ d^{\{\tq\tq\}}\end{matrix}\right]
u_{\mu}^{B}= -4 \int \frac{d^{4}l}{(2\pi)^{4}} \\
{\cal M}
\left[\begin{matrix}
d^{\{\tqu\tq\}}\\ d^{\{\tq\tq\}}\end{matrix}\right] u_{\nu}^{B} \,,
\label{eq:D2}
\end{equation}%
where
\bea
\cal{M}=\left[\begin{matrix}
{\cal M}_{\mu\nu}^{\{\tqu\tq\},\{\tqu \tq\}} & {\cal M}_{\mu\nu}^{\{\tqu \tq\},\{\tq\tq\}} \\
{\cal M}_{\mu\nu}^{\{\tq\tq\},\{\tqu \tq\}} & {\cal M}_{\mu\nu}^{\{\tq\tq\},\{\tq\tq\}}
\end{matrix}\right]
\eea
with the elements of the matrix ${\cal M}$ given by~:
\begin{equation} \nn
\begin{split}
{\cal M}_{\mu\nu}^{00}& =t^{f_{00}}\frac{1}{M_{\tqu}} \,
\Gamma_{\rho}^{1^{+}}(\ell_{\tqu\tq})  \,
\bar{\Gamma}_{\mu}^{1^{+}}(-k_{\tqu\tq}) \, S(l_{\tq}) \,
\Delta_{\rho\nu}^{1^{+}}(\ell_{\tqu\tq}), \\
{\cal M}_{\mu\nu}^{01}&=t^{f_{01}}\frac{1}{M_{\tq}} \,
\Gamma_{\rho}^{1^{+}}(\ell_{\tq\tq})  \,
\bar{\Gamma}_{\mu}^{1^{+}}(-k_{\tqu\tq}) \, S(l_{q_1}) \,
\Delta_{\rho\nu,\{\tq\tq\}}^{1^{+}}(\ell_{\tq\tq}),
\end{split}
\end{equation}
\begin{equation}
\begin{split}
{\cal M}_{\mu\nu}^{10} &=t^{f_{10}}\frac{1}{M_{\tq}} \,
\Gamma_{\rho}^{1^{+}}(\ell_{\tqu\tq})  \,
\bar{\Gamma}_{\mu}^{1^{+}}(-k_{\tq\tq}) \, S(l_{\tq}) \,
\Delta_{\rho\nu}^{1^{+}}(\ell_{\tqu\tq}), \\
{\cal M}_{\mu\nu}^{11}&=t^{f_{11}} \frac{1}{M_{\tq}} \,
\Gamma_{\rho}^{1^{+}}(\ell_{\tq\tq})  \,
\bar{\Gamma}_{\mu}^{1^{+}}(-k_{\tq\tq}) \, S(l_{\tq_1}) \,
\Delta_{\rho\nu}^{1^{+}}(\ell_{\tq\tq}),
\end{split}
\end{equation}
where $t^f$ are the flavor matrices and can be found in appendix~\ref{app:Fla}. The color-singlet bound states constructed from three heavy charm/bottom quarks are:
\begin{eqnarray}
&&\nn \begin{array}{cc}
u_{\Omega_{\tc\tc\tc}^{++*}}= \left[
\begin{array}{c}
\{\tc\tc\} \tc \\
\end{array} \right], &
\hspace{1.0cm}u_{\Omega_{\tb\tb\tb}^{-*}}=
\left[ \begin{array}{c}
\{\tb\tb\} \tb \\
\end{array} \right],
\end{array} \\ \nn \\
&&\nn \begin{array}{cc}
u_{\Omega_{\tc\tc\tb}^{+*}}= \left[
\begin{array}{c}
\{\tc\tc\}\tb \\
\{\tc\tb\} \tc \\
\end{array} \right], &
\hspace{1cm}u_{\Omega_{\tc\tb\tb}^{0*}}=
\left[ \begin{array}{c}
\{\tc\tb\} \tb \\
\{\tb\tb\} \tc
\end{array} \right],
\end{array}
\end{eqnarray}
The column vectors representing singly and doubly heavy baryons are:
\begin{eqnarray}
\hspace{-0.8cm}&&\nn \begin{array}{cc}
u_{\Sigma_{\tc}^{++*}{(\tu\tu\tc)}}=
\left[ \begin{array}{c}
\{\tu\tu\} \tc \\
\{\tu\tc\} \tu
\end{array} \right],
&
\hspace{0.3cm}u_{\Xi_{\tc\tc}^{++*}{(\tu\tc\tc)}}= \left[
\begin{array}{c}
\{\tu\tc\} \tc \\
\{\tc\tc\} \tu \\
\end{array} \right],
\end{array} \\ \nn \\
\hspace{-0.8cm}&&\nn \begin{array}{cc}
u_{\Omega_{\tc}^{0*}{(\ts\ts\tc)}}=
\left[ \begin{array}{c}
\{\ts\ts\} \tc \\
\{\ts\tc\} \ts
\end{array} \right],
 &
\hspace{0.9cm}u_{\Omega_{\tc\tc}^{+*}{(\ts\tc\tc)}}= \left[
\begin{array}{c}
\{\ts\tc\} \tc \\
\{\tc\tc\} \ts \\
\end{array} \right],
\end{array}
\\ \nn \\
\hspace{-0.8cm}&&\nn \begin{array}{cc}
u_{\Sigma_{\tb}^{+*}{(\tu\tu\tb)}}=
\left[ \begin{array}{c}
\{\tu\tu\} \tb \\
\{\tu\tb\} \tu
\end{array} \right],
&
\hspace{0.7cm}u_{\Sigma_{\tb\tb}^{0*}{(\tu\tb\tb)}}= \left[
\begin{array}{c}
\{\tu\tb\}\tb \\
\{\tb\tb\} \tu \\
\end{array} \right],
\end{array} \\ \nn \\
\hspace{-0.8cm}&&\nn \begin{array}{cc}
u_{\Omega_{\tb}^{-*}{(\ts\ts\tb)}}=
\left[ \begin{array}{c}
\{\ts\ts\} \tb \\
\{\ts\tb\} \ts
\end{array} \right],
&
\hspace{0.8cm}u_{\Omega_{\tb\tb}^{-*}{(\ts\tb\tb)}}= \left[
\begin{array}{c}
\{\ts\tb\}\tb \\
\{\tb\tb\} \ts \\
\end{array} \right],
\end{array}
\end{eqnarray}
We have solved FE~(\ref{fd1}) and
obtained masses and eigenvectors of the ground-state baryons. In case of baryons with spin $3/2$, we use $g_B=1$ for both CI-LP and CI-HP.
For triply heavy baryons, we present a comparison in Table~\ref{Table-ccc-comparison} with masses obtained using other methods.
\begin{table}[H]
\begin{center}
\caption{\label{Table-ccc-comparison} Triply heavy baryons in different approaches }
\begin{tabular}{@{\extracolsep{0.3 cm}}lllll}
\hline
     & $\Omega_{\tc\tc\tc}^{++*}$ & $\Omega_{\tb\tb\tb}^{-*}$ & $\Omega_{\tc\tc\tb}^{+*}$ & $\Omega_{\tc\tb\tb}^{0*}$  \\  CI-LP\; & $4.78$ & $14.39$ & $8.03$ & $11.10$  \\CI-HP & 4.93 & 14.23 & 8.03 & 11.12\\
     PC & $4.76$  & $14.37$ & $7.96$ & $11.17$ \\
     Lattice & 4.80 & 14.37&8.01&11.20\\
     Coulomb & 4.76 & 14.37&7.98 & 11.19\\
     Cornell &4.80 & 14.40 &8.04&11.24\\
     Fadv & 4.80 & 14.40& 8.02&11.22\\
     BM & 4.79 & 14.30&8.03&11.20\\
     QCR & 4.92 & 14.76&8.20&11.48\\
     CQM1 &4.97 & 14.83& 8.26&11.55\\
     CQM2 & 4.63&&&\\
     RQM &4.80 & 14.57& 8.02&11.29\\
     IQM &4.77&&&\\
     HCM &4.74 & 14.45& 8.10&11.38\\
     SR &4.67 & 13.28& 7.44&10.46\\
     Regee &4.82&&&\\
     NRQCD &4.90 & 14.77&8.24&11.53\\
\hline
\end{tabular}
\end{center}
\end{table}
The results with CI-LP have a difference with those obtained on lattice of less than $1\%$, while for CI-HP, the largest difference is approximately $3\%$.
%
%
The masses of the heaviest baryons with spin $3/2$,
 $\Omega_{\tc\tc\tc}$ y $\Omega_{\tb\tb\tb}$ are illustrated in Figs.~\ref{omegac1} and~\ref{omegab1}.
%

Substituting the two sets of parameters proposed in the previous sections into the FE, we now obtain the masses of singly and doubly heavy baryons. We present these results
in Table~\ref{table-baryon2} in the units of $m_{\Omega_{\tc\tc\tc}}$.
\begin{table}[H]
\begin{center}
\caption{\label{table-baryon2}
Baryon masses (in GeV). In (I) and (II) the baryon masses are listed in units of $m_{\Omega_{\tc\tc\tc}}$. The lattice values are taken from~\cite{Brown:2014ena,Mathur:2018epb} and experimental ones from~\cite{Patrignani:2016xqp,Tanabashi:2018oca} }
\begin{tabular}{@{\extracolsep{0.3 cm}}ccccc}
\hline
\hline
 (I) & $\Sigma_{\tc}^{++*}{(\tu\tu\tc)}$ & $\Xi_{\tc\tc}^{++*}{(\tu\tc\tc)}$ & $\Omega_{\tc}^{0*}{(\ts\ts\tc)}$ & $\Omega_{\tc\tc}^{+*}{(\ts\tc\tc)}$    \\
CI-LP & $0.55$ &0.77&0.60&0.79 \\
CI-HP & 0.57 & 0.79 & 0.61 & 0.82 \\
PC \; & $0.51$ & $0.75$ & $0.57$ & $0.79$   \\
Lattice & $0.52$ & $0.75$ & $0.56$ & $0.78$    \\
Expt. & 0.53 & $\cdots$ &0.58 & $\cdots$ \\
\hline
\end{tabular} \\
\medskip
\begin{tabular}{@{\extracolsep{0.3 cm}}ccccc}
\hline
   (II) & $\Sigma_{\tb}^{+*}{(\tu\tu\tb)}$ & $\Xi_{bb}^{0*}{(\tu\tb\tb)}$ & $\Omega_{\tb}^{-*}{(\ts\ts\tb)}$ & $\Omega_{\tb\tb}^{-*}{(\ts\tb\tb)}$    \\
CI-LP & $1.19$ &$2.10$&1.25&2.13 \\
CI-HP & 1.23 & 2.12 &1.28 & 2.14\\
PC & $1.18$ & $2.10$ & $1.24$ & $2.13$    \\
Lattice & $1.22$ & $2.11$ & $1.26$ & $2.14$ \\
Expt. &1.21 & $\cdots$ & $\cdots$ & $\cdots$\\
\hline
%
\hline
\end{tabular}
\end{center}
\end{table}
%
\begin{table}[htbp]
\begin{center}
\caption{\label{table-eigenvector}We list Faddeev amplitudes for spin-$3/2$ baryons.
The dominant diquarks, according to our analysis, are listed in the last column. }
\begin{tabular}{@{\extracolsep{0.3cm}}ccccc}
\hline \hline
&& $d^{\{qq_1\}}$ & $d^{\{qq\}}$ & dom. \\
 \rule{0ex}{3.0ex}
  \multirow{2}{1.5cm}{ $\Omega_{\tc\tc\tb}^{+*}$ }& CI-LP & -0.66  &-0.75 & $\{\tc\tc\}\tb$
\\ \rule{0ex}{3.0ex}
& CI-HP & -0.35 & -0.93 &$\{\tc\tc\}\tb$
\\
\hline
%
 \rule{0ex}{3.0ex}
 \multirow{2}{1.5cm}{$\Omega_{\tc\tb\tb}^{0*}$ } &CI-LP &
-0.32 &
-0.95 &
 $\{\tb\tb\}\tc$
\\ \rule{0ex}{3.0ex}
& CI-HP & -0.16 &
-0.99 & $\{\tb\tb\}\tc$ \\
\hline
%
 \rule{0ex}{3.0ex}
 \multirow{2}{1.5cm}{$\Omega_{\tc}^{0*}{(\ts\ts\tc)}$} &CI-LP &
-0.80 &
-0.60 &
 $\{\ts\tc\}\ts$
\\ \rule{0ex}{3.0ex}
& CI-HP &-0.60 &
-0.80& $\{\ts\ts\}\tc$ \\
\hline
%
 \rule{0ex}{3.0ex}
 \multirow{2}{1.5cm}{$\Omega_{\tc\tc}^{+*}{(\ts\tc\tc)}$} &CI-LP &
-0.31 &
-0.95 &
 $\{\tc\tc\}\ts$
\\ \rule{0ex}{3.0ex}
& CI-HP &-0.17 &
-0.99 & $\{\tc\tc\}\ts$ \\
\hline
\rule{0ex}{3.0ex}
 \multirow{2}{1.5cm}{$\Omega_{\tb}^{-*}{(\ts\ts\tb)}$} &CI-LP &
-0.93&
-0.36&
 $\{\ts\tb\}\ts$
\\ \rule{0ex}{3.0ex}
& CI-HP &
-0.42 &
-0.90 &
 $\{\ts\ts\}\tb$ \\
\hline
%
\rule{0ex}{3.0ex}
 \multirow{2}{1.5cm}{$\Omega_{\tb\tb}^{-*}{(\ts\tb\tb)}$} &CI-LP &
-0.11 &
-0.99 &
 $\{\tb\tb\}\ts$
\\ \rule{0ex}{3.0ex}
& CI-HP &
-0.02 &
-0.99 &
 $\{\tb\tb\}\ts$ \\
\hline
%
\rule{0ex}{3.0ex}
 \multirow{2}{1.5cm}{$\Sigma_{\tc}^{++*}{(\tu\tu\tc)}$} &CI-LP &
0.62&
0.78&
 $\{\tu\tu\}\tc$
\\ \rule{0ex}{3.0ex}
& CI-HP &
-0.67 &
-0.74 &
 $\{\tu\tu\}\tc$ \\
\hline
%
%
\rule{0ex}{3.0ex}
 \multirow{2}{1.5cm}{$\Xi_{\tc\tc}^{++*}{(\tu\tc\tc)}$} &CI-LP &
-0.26&
-0.97 &
 $\{\tc\tc\}\tu$
\\ \rule{0ex}{3.0ex}
& CI-HP &
-0.12 &
-0.99 &
 $\{\tc\tc\}\tu$ \\
\hline
\rule{0ex}{3.0ex}
 \multirow{2}{1.5cm}{$\Sigma_{\tb}^{+*}{(\tu\tu\tb)}$} &CI-LP &
-0.94 &
-0.35 &
 $\{\tu\tb\}\tb$
\\ \rule{0ex}{3.0ex}
& CI-HP &
-0.56 &
-0.83 &
 $\{\tu\tu\}\tb$ \\
\hline
%
%
\rule{0ex}{3.0ex}
 \multirow{2}{1.5cm}{$\Xi_{\tb\tb}^{0*}{(\tu\tb\tb)}$} &CI-LP &
-0.08  &
-0.99 &
 $\{\tb\tb\}\tu$
\\ \rule{0ex}{3.0ex}
& CI-HP &
0.03 &
0.99 &
 $\{\tb\tb\}\tu$ \\
\hline
\hline
\end{tabular}
\end{center}
\end{table}
The masses of baryons with spin $3/2$ with a single heavy quark obey an equal-spacing rule~\cite{Ebert:2005xj,GellMann:1962xb,Okubo:1961jc}
\bea \label{gmo-h}
m_{\Sigma_Q}+m_{\Omega_Q}=2m_{\Xi_Q} \,.\;\;\;\;\;\;\;\;\;Q=\tc,\;\tb
\eea
This relation for CI-LP yields a mass of $m_{\Xi_{\tu\ts\tc}}=2.76$ GeV while for CI-HP it is $2.83$ GeV. These predictions compare well with the experimental result of $2.467$ GeV reported in~\cite{Tanabashi:2018oca}. For the corresponding baryon containing a bottom quark, we obtain $m_{\Xi_{\tu\ts\tb}}=5.86$ GeV for CI-LP and $6.024$ GeV for CI-HP. The observation of the baryon $\Xi_{\tu\ts\tb}$ was reported by the CMS  Collaboration with a value of $5.948$ GeV~\cite{Chatrchyan:2012ni} which is in good agreement with our results.

We now turn our attention to the spacing rules which combine baryons with different
spins~\cite{Yu:2018com}~:
\bea \label{gmo-10}
\hspace{-1cm}&&m_{\Xi_{\tc\tc}^{++*}{(\tu\tc\tc)}}\hspace{-0.1cm}-\hspace{-0.09cm}m_{\Xi_{\tc\tc}^{++}(\tu\tc\tc)}\hspace{-0.1cm}-\hspace{-0.09cm}m_{\Sigma_{\tc}^{++*}{(\tu\tu\tc)}}\hspace{-0.1cm}+\hspace{-0.1cm}m_{\Sigma_{\tc}^{++}(\tu\tu\tc)}\hspace{-0.09cm}=0\,,\\
\label{gmo-11}\hspace{-1cm}&&m_{\Omega_{cc}^{+*}{(\ts\tc\tc)}}-m_{\Omega_{\tc\tc}^{+}(\ts\tc\tc)}-m_{\Omega_{\tc}^{0*}{(\ts\ts\tc)}}+m_{\Omega_{\tc}^{0}(\ts\ts\tc)}=0 \,,\\
\label{gmo-12}\hspace{-1cm}&&m_{\Xi_{\tb\tb}^{0*}{(\tu\tb\tb)}}-m_{\Xi_{\tb\tb}^{0}(\tu\tb\tb)}-m_{\Sigma_{\tb}^{+*}{(\tu\tu\tb)}}+m_{\Sigma_{\tb}^{+}(\tu\tu\tb)}=0 \,.
\eea
\begin{center}
\begin{tabular}{@{\extracolsep{0.4 cm}}cccc}
\hline \hline
Spacing rule & CI-LP & CI-HP  \\

\rule{0ex}{3.0ex}
 Eq.~(\ref{gmo-10}) &-0.01 & -0.004\\
 \rule{0ex}{3.0ex}
 Eq.~(\ref{gmo-11}) &-0.04 & 0.07\\
 \rule{0ex}{3.0ex}
 Eq.~(\ref{gmo-12}) &0.08 & -0.008 \\
 \hline \hline
\end{tabular}
\end{center}
%
We find that the CI with LP and HP parameters generates a mass spectrum that is consistent with these spacing rules. The differences of the results obtained with CI-HP and the experiment are closer to zero as compared to those obtained using CI-LP.
%
%
We define a constituent-quark passive-mass like relation~\cite{Qin:2018dqp}, via \begin{equation}
\label{EqMfP}
M_f^P = \frac{1}{3} m_{\Omega_{fff}}\,.
\end{equation}
In the following table, we compare the computed values (in GeV) from this relation
with the input parameters we used for CI-:
\bea
\label{EqMfP2}
{\rm baryon:}\quad
\begin{array}{l|cc}
\hline
\hline
f &  \tc & \tb \\
  \rule{0ex}{3.0ex}
M_f^P\; \textrm{CI-LP} & 1.59 & 4.8\\
 \rule{0ex}{3.0ex}
M_f\;\, \textrm{CI-LP}& 1.60 & 4.83 \\
\rule{0ex}{3.0ex}
M_f^P\; \textrm{CI-HP} & 1.64 & 4.74 \\
\rule{0ex}{3.0ex}
M_f\;\; \textrm{CI-HP}&1.53 &4.68 \\
\hline
\hline
\end{array}
\eea
The analogous quantity defined via ground-state vector meson masses takes very similar values (in GeV):
\begin{equation}
\label{EqMfP2meson}
{\rm meson:~}\quad
\begin{array}{l|ccc}
\hline
\hline
f &  \tc & \tb \\
 \rule{0ex}{3.0ex}
 M_f^P\; \textrm{CI-LP}  & 1.49 & 4.73\\
 \rule{0ex}{3.0ex}
 M_f\; \textrm{CI-LP}  & 1.60 & 4.83 \\
\rule{0ex}{3.0ex}
M_f^P\; \textrm{CI-HP} & 1.57 & 4.73\\
\rule{0ex}{3.0ex}
M_f\; \textrm{CI-HP} &1.53 &4.68 \\
\hline
\hline
\end{array}
\end{equation}
%
Note that the constituent-quark passive-masses
are approximately the same as our computed dressed-quark masses provided
in Table~\ref{table-M}.
%
%
\section{Conclusions}
\label{Conclusions}
The widely used CI model incorporates the key features of QCD such as chiral symmetry
breaking, confinement and low energy Golberger-Treiman relations. It has previously been employed to calculate properties of light mesons and baryons as well as heavy quarkonia. We extend the domain of applicability of this model to study the mass spectrum of heavy-light mesons, the corresponding diquarks as well as heavy baryons of positive parity. To study these systems, we use two sets of parameters: the first set consists of exactly the same parameters for all the mesons and baryons (CI-LP) while the second second is a mass fit inspired by our previous work~\cite{Raya:2017ggu} (CI- HP). Employing this fit, we propose a systematic scheme to inspect mesons in different mass range. We keep the dressed quark masses used in our previous works, and then fix the $\Lambda_{\mathrm {UV}}$ parameter in the Bethe-Salpeter equation to obtain the masses of heavy and heavy-light mesons, see Table~\ref{table-mesones-pseudo}, along with their respective coupling strength. Additionally, we calculate the masses of their diquarks partners, listed in Table~\ref{table-diquarks}, which eventually enter the computation of heavy baryon masses. A careful analysis of Tables~\ref{table-mesones-pseudo},~\ref{table-diquarks},~\ref{table-baryons-oct} and~\ref{table-baryon2} reveals that our mass predictions for both mesons and baryons are in excellent accordance with the existing experimental data.
Our results on the masses of doubly heavy baryons $(\Xi_{\tc\tc}^{++})$ agree with the ones announced by SELEX~\cite{Mattson:2002vu} and LHCb~\cite{Aaij:2017ueg}. CI with parameters presented herein provides a difference of $<1\%$ with LHCb.
There are no experimental results available for the masses of the heaviest baryons $\Omega_{\tc\tc\tc}^{++*}$ and $\Omega_{\tb\tb\tb}^{-*}$. However, our computed masses
agree fairly well with other approaches available, see Figs.~\ref{omegac1} and~\ref{omegab1}. The results illustrated in these figures show the proximity of the masses calculated in this work with the average of a plenty of predictions.
Additionally, our results for mesons and baryons satisfy the spacing rules to a high accuracy. Employing some of these rules, we can predict masses of baryons with three different flavors of quarks, Eq.~(\ref{gmo-h}).
Our study clearly shows that we can use the same parameters to calculate the masses of all mesons and baryons. However, if we want to compute other observables, such as
decay constants and form factors, we need to adapt the model better. The set of parameters CI-HP does that job for us. Note that unlike other calculations with
CI~\cite{Roberts:2011cf,Chen:2012qr,Xu:2015kta,Lu:2017cln,Yin:2019bxe}, we can fix $g_N=1$ for all baryons, still obtaining the same or better accuracy, without introducing additional parameters for any fine tuning. Our further steps of reserach will involve exotics, excited states as well form factors of
mesons and baryons containing heavy quarks.
\begin{acknowledgements}
 L. X. Guti\'errez wishes to thank the National Council for Science and Technology of Mexico (CONACyT) for the support provided to her through the programme of C\'atedras CONACyT. This research was also partly supported by Coordinaci\'on de la Investigaci\'on Cientifica (CIC) of the University of Michoacan and CONACyT, Mexico, through Grant nos. 4.10 and CB2014-22117, respectively.
\end{acknowledgements}
\appendix
\setcounter{equation}{0}
\renewcommand{\theequation}{\Alph{section}.\arabic{equation}}
\section{Euclidean Conventions}
\label{App:EM}
In our Euclidean formulation:
\begin{equation}
p\cdot q=\sum_{i=1}^4 p_i q_i\,;
\end{equation}
where
\begin{eqnarray}\nn
&&\{\gamma_\mu,\gamma_\nu\}=2\,\delta_{\mu\nu}\,;\;
\gamma_\mu^\dagger = \gamma_\mu\,;\;
\sigma_{\mu\nu}= \frac{i}{2}[\gamma_\mu,\gamma_\nu]\,; \; \\
&&{\rm tr}\,[\gamma_5\gamma_\mu\gamma_\nu\gamma_\rho\gamma_\sigma]=
-4\,\epsilon_{\mu\nu\rho\sigma}\,, \epsilon_{1234}= 1\,.
\end{eqnarray}
A positive energy spinor satisfies
\begin{equation}
\bar u(P,s)\, (i \gamma\cdot P + M) = 0 = (i\gamma\cdot P + M)\, u(P,s)\,,
\end{equation}
where $s=\pm$ is the spin label.  It is normalised as~:
\begin{equation}
\bar u(P,s) \, u(P,s) = 2 M \,,
\end{equation}
and may be expressed explicitly as~:
\begin{equation}
u(P,s) = \sqrt{M- i {\cal E}}
\left(
\begin{array}{l}
\chi_s\\
\displaystyle \frac{\vec{\sigma}\cdot \vec{P}}{M - i {\cal E}} \chi_s
\end{array}
\right)\,,
\end{equation}
with ${\cal E} = i \sqrt{\vec{P}^2 + M^2}$,
\begin{equation}
\chi_+ = \left( \begin{array}{c} 1 \\ 0  \end{array}\right)\,,\;
\chi_- = \left( \begin{array}{c} 0\\ 1  \end{array}\right)\,.
\end{equation}
For the free-particle spinor, $\bar u(P,s)= u(P,s)^\dagger \gamma_4$.
It can be used to construct a positive energy projection operator:
\begin{equation}
\label{Lplus} \Lambda_+(P):= \frac{1}{2 M}\,\sum_{s=\pm} \, u(P,s) \, \bar
u(P,s) = \frac{1}{2M} \left( -i \gamma\cdot P + M\right).
\end{equation}
A negative energy spinor satisfies
\begin{equation}
\bar v(P,s)\,(i\gamma\cdot P - M) = 0 = (i\gamma\cdot P - M) \, v(P,s)\,,
\end{equation}
and possesses properties and satisfies constraints obtained through obvious analogy
with $u(P,s)$. A charge-conjugated BS amplitude is obtained via
\begin{equation}
\label{chargec}
\bar\Gamma(k;P) = C^\dagger \, \Gamma(-k;P)^{\rm T}\,C\,,
\end{equation}
where ``T'' denotes transposing  all matrix indices and
$C=\gamma_2\gamma_4$ is the charge conjugation matrix, $C^\dagger=-C$.  Moreover, we note that
\begin{equation}
C^\dagger \gamma_\mu^{\rm T} \, C = - \gamma_\mu\,, \; [C,\gamma_5] = 0\,.
\end{equation}
We employ a Rarita-Schwinger spinor to represent a covariant spin-$3/2$ field.  The positive energy
spinor is defined by the following equations:
\begin{equation}
\label{rarita}
(i \gamma\cdot P + M)\, u_\mu(P;r) = 0\,,
\gamma_\mu u_\mu(P;r) = 0\,,
P_\mu u_\mu(P;r) = 0,
\end{equation}
where $r=-3/2,-1/2,1/2,3/2$.  It is normalised as:
\begin{equation}
\bar u_{\mu}(P;r^\prime) \, u_\mu(P;r) = 2 M\,,
\end{equation}
and satisfies a completeness relation
\begin{equation}
\label{Deltacomplete}
\frac{1}{2 M}\sum_{r=-3/2}^{3/2} u_\mu(P;r)\,\bar u_\nu(P;r) =
\Lambda_+(P)\,R_{\mu\nu}\,,
\end{equation}
where
\begin{equation}
R_{\mu\nu} = \delta_{\mu\nu} \mbox{\boldmath $I$}_{\rm D} -\frac{1}{3} \gamma_\mu \gamma_\nu +
\frac{2}{3} \hat P_\mu \hat P_\nu \mbox{\boldmath $I$}_{\rm D} - i\frac{1}{3} [ \hat P_\mu
\gamma_\nu - \hat P_\nu \gamma_\mu]\,,
\end{equation}
with $\hat P^2 = -1$. It is very useful in simplifying the FE for a positive energy decouplet state.
\vspace{-1cm}
\setcounter{equation}{0}
\section{Kernel in FE}\label{app:Fad}
\begin{align}
\nn \mathcal{M}^{11}&= t^{\tq T}t^{[\tq\tqu]}t^{[\tq\tqu]T}t^{\tq}\\
\nn &\times\{\g^{0^+}_{[\tq\tqu]}(l_{\tq\tqu}) S_{\tqu}^T\overline{\g}^{0^+}_{[\tq\tqu]}(-k_{\tq\tqu})S_{\tq}(l_{\tq})\D^{0^+}_{[\tq\tqu]}(l_{\tq\tqu})\}\\
\nn \mathcal{M}^{12}_{\nu}&= t^{\tq T}t^{\{\tq\tq\}}t^{[\tq\tqu]T}t^{\tqu} \\
\nn &\times\{\g^{1^+}_{\{\tq\tq\},\mu}(l_{\tq\tq}) S_{\tq}^T\overline{\g}^{0^+}_{[\tq\tqu]}(-k_{\tq\tqu})S_{\tqu}(l_{\tqu})\D^{1^+}_{\{\tq\tq\},\mu\nu}(l_{\tq\tq})\}\\
\nn \mathcal{M}^{13}_\nu&= t^{\tq T}t^{\{\tq\tqu\}}t^{[\tq\tqu]T}t^{\tq}\\
 \nn & \times \{\g^{1^+}_{\{\tq\tqu\},\mu}(l_{\tq\tqu}) S_{\tqu}^T\overline{\g}^{0^+}_{[\tq\tqu]}(-k_{\tq\tqu})S_{\tq}(l_{\tq})\D^{1^+}_{\{\tq\tqu\},\mu\nu}(l_{\tq\tqu})\}\\
\nn \mathcal{M}^{14}_\nu&= t^{\tq T}t^{\{\tq\tq\}}t^{[\tq\tqu]T}t^{\tqu}\\
\nn& \times \g^{1^+}_{\{\tq\tq\},\mu}(l_{\tq\tq}) S_{\tq}^T\overline{\g}^{0^+}_{[\tq\tqu]}(-k_{\tq\tqu})S_{\tqu}(l_{\tqu})\D^{1^+}_{\{\tq\tq\},\mu\nu}(l_{\tq\tq})\}\\
\nn\mathcal{M}^{15}_\nu&= t^{\tq T}t^{\{\tq\tqu\}}t^{[\tq\tqu]T}t^{\tq}\\
\nn \times & \{\g^{1^+}_{\{\tq\tqu\},\mu}(l_{\tq\tqu}) S_{\tqu}^T\overline{\g}^{0^+}_{[\tq\tqu]}(-k_{\tq\tqu})S_{\tq}(l_{\tq})\D^{1^+}_{\{\tq\tqu\},\mu\nu}(l_{\tq\tqu})\}\\
\nn \mathcal{M}^{21}_\mu&=t^{\tqu T}t^{[\tq\tqu]}t^{\{\tq\tq\}T}t^{\tq}\\
\nn\times &\{\{\g^{0^+}_{[\tq\tqu]}(l_{\tq\tqu}) S_{\tq}^T\overline{\g}^{1^+}_{\{\tq\tq\},\mu}(-k_{\tq\tq})S_{\tq}(l_{\tq})\D^{0^+}_{[\tq\tqu]}(l_{\tq\tqu})\}\\
\nn  \mathcal{M}^{22}_{\mu\nu} &=t^{\tqu T}t^{\{\tq\tq\}}t^{\{\tq\tq\}T}t^{\tqu}
\\ \times &
\nn \{\g^{1^+}_{\{\tq\tq\},\rho}(l_{\tq\tq}) S_{\tqu}^T\overline{\g}^{1^+}_{\{\tq\tq\},\mu}(-k_{\tq\tq})S_{\tqu}(l_{\tqu})\D^{1^+}_{\{\tq\tq\},\rho\nu}(l_{\tq\tq})\}\\
\nn \mathcal{M}^{23}_{\mu\nu}&=t^{\tqu T}t^{\{\tq\tqu\}}t^{\{\tq\tq\}T}t^{\tq}\\
\nn \times &\{\g^{1^+}_{\{\tq\tqu\},\rho}(l_{\tq\tqu}) S_{\tq}^T\overline{\g}^{1^+}_{\{\tq\tq\},\mu}(-k_{\tq\tq})S_{\tq}(l_{\tq})\D^{1^+}_{\{\tq\tqu\},\rho\nu}(l_{\tq\tqu})\}
\end{align}
\begin{align}
\nn \mathcal{M}^{24}_{\mu\nu} &=t^{\tqu T}t^{\{\tq\tq\}}t^{\{\tq\tq\}T}t^{\tqu}\\
\nn \times & \{\g^{1^+}_{\{\tq\tq\},\rho}(l_{\tq\tq}) S_{\tqu}^T\overline{\g}^{1^+}_{\{\tq\tq\},\mu}(-k_{\tq\tq})S_{\tqu}(l_{\tqu})\D^{1^+}_{\{\tq\tq\},\rho\nu}(l_{\tq\tq})\}\\
\nn \mathcal{M}^{25}_{\mu\nu} &=t^{\tqu T}t^{\{\tq\tqu\}}t^{\{\tq\tq\}T}t^{\tq}\\
\nn\times &\{\g^{1^+}_{\{\tq\tqu\},\rho}(l_{\tq\tqu}) S_{\tq}^T\overline{\g}^{1^+}_{\{\tq\tq\},\mu}(-k_{\tq\tq})S_{\tq}(l_{\tq})\D^{1^+}_{\{\tq\tqu\},\rho\nu}(l_{\tq\tqu})\}\\
\nn \mathcal{M}^{31}_{\mu}\ &=t^{\tq T}t^{[\tq\tqu]}t^{\{\tq\tqu\}T}t^{\tq}\\
\times &
\nn \{\g^{0^+}_{[\tq\tqu]}(l_{\tq\tqu}) S_{\tqu}^T\overline{\g}^{1^+}_{\{\tq\tqu\},\mu}(-k_{\tq\tqu})S_{\tq}(l_{\tq})\D^{0^+}_{[\tq\tqu]}(l_{\tq\tqu})\}\\
\nn \mathcal{M}^{32}_{\mu\nu}&=t^{\tq T}t^{\{\tq\tq\}}t^{\{\tq\tq\}T}t^{\tqu}\\
\nn \times
&\{\g^{1^+}_{\{\tq\tq\},\rho}(l_{\tq\tq}) S_{\tq}^T\overline{\g}^{1^+}_{\{\tq\tqu\},\mu}(-k_{\tq\tqu})S_{\tqu}(l_{\tqu})\D^{0^+}_{\{\tq\tq\},\rho\nu}(l_{\tq\tq})\}\\
\nn \mathcal{M}^{33}_{\mu\nu}&=t^{\tq T}t^{\{\tq\tqu\}}t^{\{\tq\tqu\}T}t^{\tq}\\
\nn \times & \{\g^{1^+}_{\{\tq\tqu\},\rho}(l_{\tq\tqu}) S_{\tq}^T\overline{\g}^{1^+}_{\{\tq\tqu\},\mu}(-k_{\tq\tqu})S_{\tq}(l_{\tq})\D^{0^+}_{\{\tq\tqu\},\rho\nu}(l_{\tq\tqu})\}\\
\nn \mathcal{M}^{34}_{\mu\nu}&=t^{\tq T}t^{\{\tq\tq\}}t^{\{\tq\tq\}T}t^{\tqu}\\
\nn \times & \{\g^{1^+}_{\{\tq\tq\},\rho}(l_{\tq\tq}) S_{\tq}^T\overline{\g}^{1^+}_{\{\tq\tqu\},\mu}(-k_{\tq\tqu})S_{\tqu}(l_{\tqu})\D^{0^+}_{\{\tq\tq\},\rho\nu}(l_{\tq\tq})\}\\
\nn  \mathcal{M}^{35}_{\mu\nu}&=t^{\tq T}t^{\{\tq\tqu\}}t^{\{\tq\tqu\}T}t^{\tq}\\
\nn \times & \{\g^{1^+}_{\{\tq\tqu\},\rho}(l_{\tq\tqu}) S_{\tq}^T\overline{\g}^{1^+}_{\{\tq\tqu\},\mu}(-k_{\tq\tqu})S_{\tq}(l_{\tq})\D^{0^+}_{\{\tq\tqu\},\rho\nu}(l_{\tq\tqu})\}\\
\nn \mathcal{M}^{41}_\mu&=t^{\tqu T}t^{[\tq\tqu]}t^{\{\tq\tq\}T}t^{\tq}\\
\nn \times &\{\{\g^{0^+}_{[\tq\tqu]}(l_{\tq\tqu}) S_{\tq}^T\overline{\g}^{1^+}_{\{\tq\tq\},\mu}(-k_{\tq\tq})S_{\tq}(l_{\tq})\D^{0^+}_{[\tq\tqu]}(l_{\tq\tqu})\}\\
\nn  \mathcal{M}^{42}_{\mu\nu}&=t^{\tqu T}t^{\{\tq\tq\}}t^{\{\tq\tq\}T}t^{\tqu}\\
\nn \times & \{\g^{1^+}_{\{\tq\tq\},\rho}(l_{\tq\tq}) S_{\tqu}^T\overline{\g}^{1^+}_{\{\tq\tq\},\mu}(-k_{\tq\tq})S_{\tqu}(l_{\tqu})\D^{1^+}_{\{\tq\tq\},\rho\nu}(l_{\tq\tq})\}\\
\nn \mathcal{M}^{43}_{\mu\nu}&=t^{\tqu T}t^{\{\tq\tqu\}}t^{\{\tq\tq\}T}t^{\tq}\\
\nn & \times\{\g^{1^+}_{\{\tq\tqu\},\rho}(l_{\tq\tq}) S_{\tq}^T\overline{\g}^{1^+}_{\{\tq\tq\},\mu}(-k_{\tq\tq})S_{\tq}(l_{\tq})\D^{1^+}_{\{\tq\tqu\},\rho\nu}(l_{\tq\tqu})\}\\
\nn  \mathcal{M}^{44}_{\mu\nu} &=t^{\tqu T}t^{\{\tq\tq\}}t^{\{\tq\tq\}T}t^{\tqu} \\
\nn \times & \{\g^{1^+}_{\{\tq\tq\},\rho}(l_{\tq\tq}) S_{\tqu}^T\overline{\g}^{1^+}_{\{\tq\tq\},\mu}(-k_{\tq\tq})S_{\tqu}(l_{\tqu})\D^{1^+}_{\{\tq\tq\},\rho\nu}(l_{\tq\tq})\}\\
\nn \mathcal{M}^{45}_{\mu\nu}&=t^{\tqu T}t^{\{\tq\tqu\}}t^{\{\tq\tq\}T}t^{\tq} \\
\nn
\times & \{\g^{1^+}_{\{\tq\tqu\},\rho}(l_{\tq\tqu}) S_{\tq}^T\overline{\g}^{1^+}_{\{\tq\tq\},\mu}(-k_{\tq\tq})S_{\tq}(l_{\tqu})\D^{1^+}_{\{\tq\tqu\},\rho\nu}(l_{\tq\tqu})\}\\
\nn  \mathcal{M}^{51}_{\mu}&=t^{\tq T}t^{[\tq\tqu]}t^{\{\tq\tqu\}T}t^{\tq}\\
\nn \times & \{\g^{0^+}_{[\tq\tqu]}(l_{\tq\tqu}) S_{\tqu}^T\overline{\g}^{1^+}_{\{\tq\tqu\},\mu}(-k_{\tq\tqu})S_{\tq}(l_{\tq})\D^{0^+}_{[\tq\tqu]}(l_{\tq\tqu})\}\\
\nn \mathcal{M}^{52}_{\mu\nu}&=t^{\tq T}t^{\{\tq\tq\}}t^{\{\tq\tq\}T}t^{\tqu}\\
\nn \times & \{\g^{1^+}_{\{\tq\tq\},\rho}(l_{\tq\tq}) S_{\tq}^T\overline{\g}^{1^+}_{\{\tq\tqu\},\mu}(-k_{\tq\tqu})S_{\tqu}(l_{\tqu})\D^{0^+}_{\{\tq\tq\},\rho\nu}(l_{\tq\tq})\}\\
\nn \mathcal{M}^{53}_{\mu\nu}&=t^{\tq T}t^{\{\tq\tqu\}}t^{\{\tq\tqu\}T}t^{\tq}\\
\nn \times &\{\g^{1^+}_{\{\tq\tqu\},\rho}(l_{\tq\tqu}) S_{\tq}^T\overline{\g}^{1^+}_{\{\tq\tqu\},\mu}(-k_{\tq\tqu})S_{\tq}(l_{\tq})\D^{0^+}_{\{\tq\tqu\},\rho\nu}(l_{\tq\tqu})\}\\
\nn \mathcal{M}^{54}_{\mu\nu}&=t^{\tq T}t^{\{\tq\tq\}}t^{\{\tq\tq\}T}t^{\tqu}
 \\ \nn \times & \{\g^{1^+}_{\{\tq\tq\},\rho}(l_{\tq\tq}) S_{\tq}^T\overline{\g}^{1^+}_{\{\tq\tqu\},\mu}(-k_{\tq\tqu})S_{\tqu}(l_{\tqu})\D^{0^+}_{\{\tq\tq\},\rho\nu}(l_{\tq\tq})\}\\
\nn \mathcal{M}^{55}_{\mu\nu}&=t^{\tq T}t^{\{\tq\tqu\}}t^{\{\tq\tqu\}T}t^{\tq}
\\ \nn \times &\{\g^{1^+}_{\{\tq\tqu\},\rho}(l_{\tq\tqu}) S_{\tq}^T\overline{\g}^{1^+}_{\{\tq\tqu\},\mu}(-k_{\tq\tqu})S_{\tq}(l_{\tq})\D^{0^+}_{\{\tq\tqu\},\rho\nu}(l_{\tq\tqu})\}\\
\end{align}

The S-meson and AV-diquark BS amplitudes assume the simple form in Eqs.~(\ref{LBSEI},\ref{mes-vec}). $\Delta^{0^+}(\ell_{qq})$ and $\Delta_{\mu\nu}^{1^+}(\ell_{qq}) $, are standard
 propagators for scalar and vector diquarks.
 \begin{eqnarray}
\Delta^{0^+}(K)&=&\frac{1}{K^2+m_{qq}^2},\\
\Delta^{1^+}_{\mu\nu}(K)&=&\frac{1}{K^2+m_{qq}^2}\left(\delta_{\mu\nu}+\frac{K_\mu K_\nu}{m_{qq}^2}\right).
\end{eqnarray}
\section{Flavor Diquarks}
\label{app:Fla}
We define the following set of flavor column matrices,
\begin{equation}\nn
t^{\tu}=\begin{pmatrix} 1  \\ 0  \\ 0 \\ 0 \\0 \end{pmatrix},\;\;\;\;
t^{\td}=\begin{pmatrix} 0  \\ 1  \\ 0  \\ 0\\0  \end{pmatrix},\;\;\;\;
t^{\ts}=\begin{pmatrix} 0  \\ 0  \\ 1 \\ 0 \\0  \end{pmatrix},\;\;\;\;
\end{equation}
\begin{equation}
t^{\tc}=\begin{pmatrix} 0  \\ 0  \\ 0 \\ 1\\0  \end{pmatrix},\;\;\;\;
t^{\tb}=\begin{pmatrix} 0  \\ 0  \\ 0 \\ 0\\1  \end{pmatrix},\;\;\;\;
\end{equation}
and
\bea \nn
t^f=\left[\begin{matrix} \nn
t^{\tq T}t^{\{\tqu \tq\}}t^{\{\tqu\tq\}T}t^{\tq} &&t^{\tq T}t^{\{\tq\tq\}}t^{\{\tqu\tq\}T}t^{\tqu}\\ \\
t^{\tqu T}t^{\{\tqu\tq\}}t^{\{\tq\tq\}T}t^{\tq} && t^{\tqu T}t^{\{\tq\tq\}}t^{\{\tq\tq\}T}t^{\tqu}
\end{matrix}\right] \,.
\eea
The flavor matrices for the diquarks are
\begin{equation}\nonumber
\begin{array}{cc}
\bf{t^{[\textcolor{red}{u}\textcolor{blue}{d}]}=\begin{pmatrix} 0 & 1 & 0 & 0 & 0 \\ -1 & 0 & 0  & 0& 0 \\ 0 &  0 & 0  & 0 & 0 \\ 0 &  0 & 0  & 0 & 0\\  0 & 0 & 0 & 0 & 0 \end{pmatrix}}
&
\bf{t^{[\textcolor{red}{u}\textcolor{green}{s}]}=\begin{pmatrix} 0 & 0 & 1& 0 & 0  \\ 0 & 0 & 0& 0& 0 \\ -1 &  0 & 0& 0 & 0\\ 0 &  0 & 0  & 0& 0\\  0 & 0 & 0 & 0 & 0 \end{pmatrix}}
\\[10ex]
\bf{t^{[\textcolor{blue}{d}\textcolor{green}{s}]}=\begin{pmatrix} 0 & 0 & 0 & 0 & 0 \\ 0 & 0 & 1& 0 & 0  \\ 0 &  -1 & 0 & 0& 0   \\ 0 &  0 & 0  & 0& 0 \\  0 & 0 & 0 & 0 & 0 \end{pmatrix}}
&
\bf{t^{[\tu\tc]}=\begin{pmatrix} 0 & 0 & 0 & 1& 0  \\ 0 & 0 & 0& 0& 0   \\ 0 &  0 & 0 & 0& 0   \\ -1 &  0 & 0  & 0& 0\\  0 & 0 & 0 & 0 & 0 \end{pmatrix}},
\\[10ex]
\bf{t^{\{\textcolor{red}{u}\textcolor{red}{u}\}}=\begin{pmatrix} \sqrt{2} & 0 & 0& 0 &0\\ 0 & 0 & 0& 0&0 \\ 0 &  0 & 0 & 0&0 \\ 0 &  0 & 0 & 0&0\\ 0 & 0 & 0& 0& 0  \end{pmatrix}}
&
\bf{t^{\{\textcolor{red}{u}\textcolor{blue}{d}\}}=\begin{pmatrix} 0 & 1 & 0 & 0&0\\ 1 & 0 & 0& 0&0 \\ 0 &  0 & 0& 0&0 \\ 0 &  0 & 0 & 0 &0\\ 0 & 0 & 0& 0& 0  \end{pmatrix}},
\\[10ex]
\end{array}
\end{equation}
\begin{equation}\nonumber
\begin{array}{cc}
\bf{t^{\{\tu\ts\}}=\begin{pmatrix} 0 & 0 & 1& 0&0  \\ 0 & 0 & 0& 0&0  \\ 1 &  0 & 0& 0 &0 \\ 0 &  0 & 0 & 0&0 \\ 0 & 0 & 0& 0& 0   \end{pmatrix}}
&
t^{\{\tu\tc\}}=\begin{pmatrix} 0 & 0 & 0& 1&0 \\ 0 & 0 & 0 & 0&0\\ 0 &  0 & 0 & 0&0\\  1 &  0 & 0 & 0&0\\ 0 & 0 & 0& 0& 0  \end{pmatrix},
\\[10ex]
\bf{t^{\{\td\td\}}=\begin{pmatrix} 0 & 0 & 0 & 0 & 0 \\ 0 &  \sqrt{2} & 0& 0 & 0 \\ 0 &  0 & 0& 0& 0  \\ 0 & 0 & 0 & 0& 0\\ 0 & 0 & 0& 0& 0  \end{pmatrix}}
&
\bf{t^{\{\td\ts\}}=\begin{pmatrix} 0 & 0 & 0& 0 & 0 \\ 0 & 0 & 1& 0 & 0\\ 0 &  1 & 0& 0 & 0\\ 0 & 0 & 0 & 0& 0\\ 0 & 0 & 0& 0& 0   \end{pmatrix}},\;\;\;\;
\\[10ex]
\bf{t^{\{\ts\ts\}}=\begin{pmatrix} 0 & 0 & 0& 0& 0 \\ 0 & 0 & 0& 0& 0 \\ 0 &  0 & \sqrt{2} & 0 & 0 \\ 0 & 0 & 0 & 0 & 0\\ 0 & 0 & 0& 0& 0  \end{pmatrix}},
&
t^{\{\tc\tc\}}=\begin{pmatrix} 0 & 0 & 0& 0 & 0 \\ 0 & 0 & 0 & 0 & 0\\ 0 &  0 & 0 & 0 & 0\\  0 &  0 & 0 &  \sqrt{2} & 0\\ 0 & 0 & 0& 0& 0   \end{pmatrix}\\[10ex]
\bf{t^{[\td\tc]}}=\begin{pmatrix} 0 & 0 & 0& 0& 0 \\ 0 & 0 & 0 & 1& 0\\ 0 &  0 & 0 & 0& 0\\  0 &  -1& 0 & 0& 0 \\ 0 & 0 & 0& 0& 0    \end{pmatrix},
&
\bf{t^{[\ts\tc]}}=\begin{pmatrix} 0 & 0 & 0& 0 & 0 \\ 0 & 0 & 0 & 0& 0\\ 0 &  0 & 0 & 1& 0\\  0 &  0& -1 & 0& 0 \\ 0 & 0 & 0& 0& 0    \end{pmatrix},\\[10ex]
%
\bf{t^{\{\td\tc\}}}=\begin{pmatrix} 0 & 0 & 0& 0 & 0\\ 0 & 0 & 0 & 1& 0\\ 0 &  0 & 0 & 0& 0\\  0 &  1& 0 & 0& 0 \\ 0 & 0 & 0& 0& 0   \end{pmatrix},
&
\bf{t^{\{\ts\tc\}}}=\begin{pmatrix} 0 & 0 & 0& 0 & 0 \\ 0 & 0 & 0 & 0 & 0\\ 0 &  0 & 0 & 1 & 0\\  0 &  0& 1 & 0 & 0 \\ 0 & 0 & 0& 0& 0   \end{pmatrix},
\end{array}
\end{equation}
\begin{equation}\nonumber
\begin{array}{cc}
\bf{t^{[\tb\tu]}}=\begin{pmatrix} 0 & 0 & 0& 0& -1\\ 0 & 0 & 0 & 0& 0\\ 0 &  0 & 0 & 0& 0\\  0 &  0& 0 & 0& 0 \\ 1 & 0 & 0& 0& 0    \end{pmatrix},
&
\bf{t^{[\tb\td]}}=\begin{pmatrix} 0 & 0 & 0& 0 & 0 \\ 0 & 0 & 0 & 0& -1\\ 0 &  0 & 0 & 0& 0\\  0 &  0& 0 & 0& 0 \\ 0 & 1 & 0& 0& 0    \end{pmatrix},
\\[10ex]
\bf{t^{[\tb\tc]}}=\begin{pmatrix} 0 & 0 & 0& 0 & 0 \\ 0 & 0 & 0 & 0& 0\\ 0 &  0 & 0 & 0& 0\\  0 &  0& 0 & 0& -1 \\ 0 & 0 & 0& 1& 0    \end{pmatrix},
&
\bf{t^{[\tb\ts]}}=\begin{pmatrix} 0 & 0 & 0& 0 & 0 \\ 0 & 0 & 0 & 0& 0\\ 0 &  0 & 0 & 0& -1\\  0 &  0& 0 & 0& 0 \\ 0 & 0 & 1& 0& 0    \end{pmatrix},\;\;\;\;
\\[10ex]
\bf{t^{\{\tb\tb\}}}=\begin{pmatrix} 0 & 0 & 0& 0& 0\\ 0 & 0 & 0 & 0& 0\\ 0 &  0 & 0 & 0& 0\\  0 &  0& 0 & 0& 0 \\ 0 & 0 & 0& 0&  \sqrt{2}    \end{pmatrix}
&
\bf{t^{\{\tb\tu\}}}=\begin{pmatrix} 0 & 0 & 0& 0& 1\\ 0 & 0 & 0 & 0& 0\\ 0 &  0 & 0 & 0& 0\\  0 &  0& 0 & 0& 0 \\ 1 & 0 & 0& 0& 0 \end{pmatrix},\;\;\;\;
\\[10ex]
\bf{t^{\{\tb\td\}}}=\begin{pmatrix} 0 & 0 & 0& 0& 0\\ 0 & 0 & 0 & 0& 1\\ 0 &  0 & 0 & 0& 0\\  0 &  0& 0 & 0& 0 \\ 0 & 1 & 0& 0& 0   \end{pmatrix}
&
\bf{t^{\{\tb\tc\}}}=\begin{pmatrix} 0 & 0 & 0& 0& 0\\ 0 & 0 & 0 & 0& 0\\ 0 &  0 & 0 & 0& 0\\  0 &  0& 0 & 0& 1 \\ 0 & 0 & 0& 1& 0   \end{pmatrix},
\\[10ex]
\bf{t^{\{\tb\ts\}}}=\begin{pmatrix} 0 & 0 & 0& 0& 0\\ 0 & 0 & 0 & 0& 0\\ 0 &  0 & 0 & 0& 1\\  0 &  0& 0 & 0& 0 \\ 0 & 0 & 1& 0& 0
&
 \end{pmatrix}.
 \end{array}
\end{equation}


\end{document}